\documentclass[%
aip,
 amsmath,amssymb,
 reprint,%
]{revtex4-1}

\usepackage{graphicx}
\usepackage{dcolumn}
\usepackage{bm}

\usepackage[utf8]{inputenc}
\usepackage[T1]{fontenc}
\usepackage{mathptmx}
\usepackage{color}

\begin{document}

\preprint{AIP/123-QED}


\title[ ]{Impact of band-anticrossing on band-to-band tunneling in highly-mismatched semiconductor alloys}

\author{Sarita Das}
 \affiliation{Tyndall National Institute, University College Cork, Lee Maltings, Dyke Parade, Cork T12 R5CP, Ireland}
 \affiliation{Department of Physics, University College Cork, Cork T12 YN60, Ireland}

\author{Christopher A.~Broderick}
 \email{c.broderick@umail.ucc.ie}
 \affiliation{Tyndall National Institute, University College Cork, Lee Maltings, Dyke Parade, Cork T12 R5CP, Ireland}
 \affiliation{Department of Physics, University College Cork, Cork T12 YN60, Ireland}

\author{Eoin P.~O'Reilly}
 \affiliation{Tyndall National Institute, University College Cork, Lee Maltings, Dyke Parade, Cork T12 R5CP, Ireland}
 \affiliation{Department of Physics, University College Cork, Cork T12 YN60, Ireland}

\date{\today}


\begin{abstract}

We theoretically analyse band-to-band tunneling (BTBT) in highly-mismatched, narrow-gap dilute nitride and bismide alloys, and quantify the impact of the N- or Bi-induced perturbation of the band structure -- due to band-anticrossing (BAC) with localised impurity states -- on the electric field-dependent BTBT generation rate. For this class of semiconductors the assumptions underpinning the widely-employed Kane model of direct BTBT break down, due to the strong band edge nonparabolicity resulting from BAC interactions. Via numerical calculations based on the Wentzel-Kramers-Brillouin approximation we demonstrate that BAC leads, at fixed band gap, to reduced (increased) BTBT current at low (high) applied electric fields compared to that in a conventional InAs$_{1-x}$Sb$_{x}$ alloy. Our analysis reveals that BTBT in InN$_{x}$As$_{1-x}$ and InAs$_{1-x}$Bi$_{x}$ is governed by a field-dependent competition between the impact of N (Bi) incorporation on (i) the dispersion of the complex band linking the valence and conduction bands, which dominates at low field strengths, and (ii) the conduction (valence) band edge density of states, which dominates at high field strengths. The implications of our results for applications in avalanche photodiodes and tunneling field-effect transistors are discussed.

\end{abstract}

\maketitle


\section{Introduction}
\label{sec:introduction}


Over the past two decades highly-mismatched III-V semiconductor alloys containing nitrogen (N) and bismuth (Bi) have attracted significant research interest due to their unusual electronic properties. The impurity-like behaviour of substitutional N or Bi atoms in conventional III-V semiconductors strongly perturbs the electronic structure, providing significant opportunities for band structure engineering to obtain properties suitable for applications in a range of devices. \cite{Broderick_nitride_chapter_2017,Broderick_bismide_chapter_2017} Specifically, due to their large band gap reduction and modified band dispersion, near- and mid-infrared semiconductor lasers, \cite{Broderick_SST_2012,Broderick_IEEEJSTQE_2015,Gladysiewicz_JAP_2015,Hader_SPIE_2016,Marko_IEEEJSTQE_2017,Delorme_APL_2017,Broderick_SST_2018,Arkani_IEEENANO_2018} as well as highly-efficient multi-junction solar cells, \cite{Jackrel_JAP_2007,Wiemer_PSPIE_2011,Sabnis_AIPCP_2012,Miyashita_PP_2016} based on dilute nitride and bismide alloys have been proposed and demonstrated, while additional work has highlighted the potential of highly-mismatched alloys for applications in long-wavelength photodetectors \cite{Webster_JAP_2016} and avalanche photodiodes (APDs), \cite{Adams_EL_2004} in intermediate-band solar cells, \cite{Kuang_APL_2013,Kudrawiec_PRA_2014} and in tunable low-frequency plasmonics. \cite{Allami_PRB_2021}


Here, we address an as yet unexplored aspect of the electronic properties of dilute nitride and bismide alloys: the complex band structure and its consequences for band-to-band tunneling (BTBT). Dependent on the specific device application of interest, BTBT can be either a desirable enabling property or a deleterious loss mechanism. For example, BTBT is the physical mechanism enabling the operation of tunneling field-effect transistors (TFETs), \cite{Ionescu_Nature_2011,Avci_IEEEJEDS_2015,Lind_IEEEJDS_2015,Convertino_JPCM_2018} making enhancement of the BTBT generation rate at fixed band gap and applied electric field desirable. However, BTBT can contribute significantly to leakage currents in long-wavelength APDs, \cite{Mahmoodi_IEEE_2015,Campbell_JLT_2016,Sweeney_chapter_2017} making suppression of the BTBT generation rate desirable for the realisation of detectors displaying high signal-to-noise ratio. \cite{Yi_NP_2019,Jones_NP_2020} Given the broad scope for applications of TFETs as an enabling platform for post-complementary metal-oxide semiconductor electronics, and of APDs for sensing at application-rich near- and mid-infrared wavelengths, it is of both fundamental and practical interest to investigate novel approaches to engineer BTBT in semiconductors.


Since BTBT is maximised in narrow-gap materials we focus on InN$_{x}$As$_{1-x}$ and InAs$_{1-x}$Bi$_{x}$ alloys, due to the strong BTBT in the direct-gap host matrix semiconductor InAs. It is well established that the main features of the band structure of dilute nitride and bismide alloys can be described via the band-anticrossing (BAC) model, \cite{Shan_PRL_1999,Shan_PSSB_2001,Wu_SST_2002} in which the extended Bloch states of the host matrix semiconductor -- the conduction band (CB) edge in dilute nitrides, \cite{Shan_PRL_1999,Lindsay_SSC_1999,Lindsay_SSE_2003} and the valence band (VB) edge in dilute bismides \cite{Alberi_PRB_2007,Alberi_APL_2007,Usman_PRB_2011} -- interact with and repel in energy the dispersionless localised impurity level associated with an isovalent substitutional N or Bi impurity. In this manner the BAC model describes (i) the strong reduction and composition-dependent bowing of the band gap due to N or Bi incorporation, and (ii) the increase in CB (VB) edge effective mass due to hybridisation of an extended band edge state with a N- (Bi-) related localised impurity state. \cite{Lindsay_SSC_1999,Lindsay_PRL_2004,Reilly_SST_2009}

Conventional analysis of direct ($\Delta \textbf{k} = 0$) BTBT based on, e.g., the analytical Kane model \cite{Kane_JPCS_1959,Kane_JAP_1961} and several popular modifications thereof, \cite{Pan_JAP_2014,Carrillo-Nunez_JAP_2015} implicitly assume parabolic CB and VB edge dispersion. Such models are therefore ill-suited to analyse BTBT in highly-mismatched alloys. As such, to allow explicit treatment of the strongly perturbed complex band structure, our analysis employs direct numerical calculation of the BTBT transmission coefficient and generation rate. To achieve this, we begin with model \textbf{k}$\cdot$\textbf{p} Hamiltonians which include N- or Bi-related BAC interactions and are parametrised directly via atomistic alloy supercell electronic structure calculations. We calculate analytically the complex band structure admitted by these model Hamiltonians, with these perturbed complex band structures then used to compute the BTBT transmission coefficient, and hence BTBT generation rate, numerically in the Wentzel-Kramers-Brillouin (WKB) approximation. To quantify the impact of N- or Bi-related BAC we compare the results of our calculations for highly-mismatched InN$_{x}$As$_{1-x}$ and InAs$_{1-x}$Bi$_{x}$ to equivalent calculations for the conventional III-V alloy InAs$_{1-x}$Sb$_{x}$.

In the semi-classical WKB approximation the BTBT transmission coefficient is a strong (exponential) function of the area bounded by the complex band linking the CB and VB in the plane defined by a carrier's energy and the imaginary component of its wave vector along the tunneling direction. This area, and hence the BTBT transmission coefficient and generation rate, depends critically on the band gap and the CB and VB edge effective masses, which respectively determine the extent of this complex band in energy and in wave vector. The strong modification of these properties in response to N or Bi incorporation suggests significant implications for BTBT -- where, at fixed band gap, the increase in band edge effective mass is expected to strongly decrease the generation rate -- and hence the potential to engineer the generation rate for device applications. Our analysis identifies that BAC-induced modification of the CB structure (in InN$_{x}$As$_{1-x}$) or VB structure (in InAs$_{1-x}$Bi$_{x}$) strongly impacts the BTBT generation rate. We demonstrate that, at fixed band gap, this impact is manifested via a field-dependent competition originating in the increased band edge effective mass: at low applied electric field $F \lesssim 1$ MV cm$^{-1}$ the increase in effective mass increases the area bounded by the complex band, reducing the generation rate, while at high applied electric field $F \gtrsim 1$ MV cm$^{-1}$ the increased density of states (DOS) makes more states available to contribute to BTBT in a given energy range, increasing the generation rate. As such, in InN$_{x}$As$_{1-x}$ and InAs$_{1-x}$Bi$_{x}$ we find that BAC leads to reduced BTBT generation rate at low applied electric field compared to that in a conventional InAs$_{1-x}$Sb$_{x}$ alloy having the same band gap, due to the increase in band edge effective mass. Conversely, at high applied electric field we find that that BAC leads to increased BTBT generation rate, due to contributions to BTBT from states having larger wave vectors in the plane perpendicular to the tunneling direction. Repeating this analysis as a function of alloy band gap, we find that the relative BTBT generation rate in InN$_{x}$As$_{1-x}$ (InAs$_{1-x}$Bi$_{x}$) can be reduced by up to $\approx 20$\% ($\approx 50$\%) compared to that in InAs$_{1-x}$Sb$_{x}$ at low applied field, while at high applied field it exceeds that in InAs$_{1-x}$Sb$_{x}$ by an amount dependent on the BAC-induced increase in the DOS close in energy to the N- or Bi-related localised impurity state in InN$_{x}$As$_{1-x}$ or InAs$_{1-x}$Bi$_{x}$ respectively. On this basis, we comment on the implications of our results for potential applications of narrow-gap highly-mismatched alloys in TFETs and in long-wavelength APDs.


The remainder of this paper is organised as follows. In Sec.~\ref{sec:theory} we describe our theoretical model, focusing in Sec.~\ref{sec:theory_hamiltonians} on model Hamiltonians, and in Sec.~\ref{sec:theory_btbt} on calculation of the BTBT transmission coefficient and generation rate. Our results are presented in Sec.~\ref{sec:results}, beginning in Sec.~\ref{sec:results_InAs} with comparison between the BTBT transmission coefficient and generation rate of InAs calculated via the analytical Kane model and via numerical WKB calculations. In Sec.~\ref{sec:results_fixed_band_gap} we compare the complex band structure and field-dependent transmission coefficient and generation rate in conventional and highly-mismatched alloys having equal band gap. Next, in Sec.~\ref{sec:results_variable_band_gap} we describe trends in the field-dependent BTBT generation rate as a function of alloy band gap. Then, in Sec.~\ref{sec:implications} we describe the implications of our results for device applications. Finally, in Sec.~\ref{sec:conclusions} we summarise and conclude.


\section{Theoretical model}
\label{sec:theory}

There exist numerous methods to compute semiconductor complex band structure, ranging from continuum multi-band \textbf{k}$\cdot$\textbf{p} Hamiltonians to atomistic tight-binding, empirical pseudopotential and first principles approaches. Similarly, given the complex band structure there are several different approaches by which to compute the BTBT transmission coefficient and generation rate, including the semi-classical WKB approximation and more sophisticated quantum kinetic approaches based on non-equilibrium Green's functions \cite{Luisier_JAP_2010,Carrillo-Nunez_JAP_2015,Luisier_PRB_2006} or Wigner functions. \cite{Yamada_IEEETED_2009,Morandi_PRB_2009} Previous analysis has demonstrated that an appropriately parametrised \textbf{k}$\cdot$\textbf{p} Hamiltonian employed in conjunction with the WKB approximation captures the key qualitative (and, for the most part, quantitative) physics of direct BTBT compared to full quantum kinetic calculations, \cite{Pan_JAP_2014} while providing the benefits of theoretical simplicity and physical transparency.

As we are interested here in analysing the impact of BAC on BTBT resulting from modification of the complex band structure, it is simplest to employ a combined \textbf{k}$\cdot$\textbf{p} and WKB approach to do so. While the WKB approximation has known quantitative shortcomings -- e.g.~underestimation of the low-field generation rate for larger band gaps \cite{Pan_JAP_2014} -- we focus on the relative impact of BAC on BTBT by comparing results for highly-mismatched InN$_{x}$As$_{1-x}$ and InAs$_{1-x}$Bi$_{x}$ to those for conventional InAs$_{1-x}$Sb$_{x}$ alloys having equal band gap. In this manner, our analysis provides detailed insight into the impact of BAC on BTBT, which is rooted in the complex band structure, with the implications for BTBT identified via relative analysis expected to hold in more sophisticated formalisms. We therefore begin with a simple 2-band \textbf{k}$\cdot$\textbf{p} Hamiltonian for the band structure of InAs, which is then modified to an appropriate 3-band BAC Hamiltonian for InN$_{x}$As$_{1-x}$ or InAs$_{1-x}$Bi$_{x}$. This allows for analytical calculation of the complex band structure associated with the BAC model, while our numerical WKB calculations produce results indistinguishable from the Kane model \cite{Kane_JPCS_1959,Kane_JAP_1961} for InAs (cf.~Sec.~\ref{sec:results_InAs}).


\subsection{Model Hamiltonians and complex band structure}
\label{sec:theory_hamiltonians}

Our description of the InAs$_{1-x}$Sb$_{x}$ alloy band structure is based on the 2-band \textbf{k}$\cdot$\textbf{p} Hamiltonian due to Kane, \cite{Kane_JPCS_1957,Kane_SS_1966} which describes the dispersion of the lowest energy conduction band (CB) and the light-hole (LH) VB in a direct-gap zinc blende semiconductor as

\begin{equation}
	H = \left( \begin{array}{cc}
	E_{g} + \frac{ \hbar^{2} }{ 2 m_{0} } \left( k_{\perp}^{2} + k_{z}^{2} \right) & i P \, \sqrt{ k_{\perp}^{2} + k_{z}^{2} }                              \\
 	-i P \, \sqrt{ k_{\perp}^{2} + k_{z}^{2} }                                     & \frac{ \hbar^{2} }{ 2 m_{0} } \left( k_{\perp}^{2} + k_{z}^{2} \right) \\
     \end{array} \right) \, ,
    \label{eq:2_band_hamiltonian}
 \end{equation}

\noindent
where $E_{g}$ is the zone-centre ($k = 0$) band gap, and $P$ is the inter-band (Kane) momentum matrix element. While the band dispersion admitted by Eq.~\eqref{eq:2_band_hamiltonian} is spherical -- i.e.~it depends only on the magnitude $k = \sqrt{ k_{\perp}^{2} + k_{z}^{2} }$ of the wave vector -- in order to treat BTBT in the presence of an electric field applied along the [001] direction we have, for clarity, expressed the wave vector $k$ in terms of its components parallel ($k_{z}$) and perpendicular ($k_{\perp}$) to [001].

The CB and LH dispersion is obtained straightforwardly by computing the roots of the characteristic equation associated with Eq.~\eqref{eq:2_band_hamiltonian}

\textcolor{black}{\begin{equation}
    E_{\pm} ( k ) = \frac{ E_{g} }{ 2 } + \frac{ \hbar^{2} k^{2} }{ 2 m_{0} } \pm \sqrt{ \left( \frac{ E_{g} }{ 2 } \right)^{2} + E_{P} \, \frac{ \hbar^{2} k^{2} }{ 2 m_{0} } } \, ,
    \label{eq:2_band_real_bands}
\end{equation}}

\noindent
where the $+$ and $-$ signs respectively describe the CB and LH dispersion, and $E_{P} = \frac{ 2 m_{0} P^{2} }{ \hbar^{2} }$ is the Kane parameter.

To obtain a description of the band structure suitable to compute the BTBT generation rate, it is critical that the CB and LH band edge effective masses $m_{\pm}^{\ast}$ are appropriately parametrised, so that the curvature of the complex band linking the VB and CB is accurately described. We therefore proceed by using Eq.~\eqref{eq:2_band_real_bands} to compute the zone-centre CB and LH effective masses

\begin{equation}
    \frac{ m^{\ast}_{\pm} }{ m_{0} } = \left| \frac{ E_{g} }{ E_{g} \pm E_{P} } \right| \, ,
    \label{eq:2_band effective_masses}
\end{equation}

\noindent
from which we note that the band edge effective masses are parametrisable via the choice of Kane parameter $E_{P}$. For InAs$_{1-x}$Sb$_{x}$ alloys we begin with the InAs room temperature band gap $E_{g} = 0.354$ eV, \cite{Vurgaftman_JAP_2001} and adjust $E_{P}$ to fit to the InAs CB edge effective mass $m_{+}^{\ast} = 0.026 \, m_{0}$, \cite{Vurgaftman_JAP_2001} yielding $E_{P} = 13.261$ eV ($P = 7.108$ eV \AA). Repeating this procedure for InSb, with $E_{g} = 0.174$ eV and $m_{+}^{\ast} = 0.0135 \, m_{0}$, \cite{Vurgaftman_JAP_2001} we obtain $E_{P} = 12.641$ eV ($P = 6.940$ eV \AA). For InAs$_{1-x}$Sb$_{x}$ alloys, we obtain the inter-band momentum matrix element $P$ via linear interpolation of the values determined for InAs and InSb.

Using this procedure we have determined $P$ by fitting solely to the CB edge effective mass $m_{+}^{\ast}$, which then fixes the LH effective mass $m_{-}^{\ast}$ (cf.~Eq.~\eqref{eq:2_band effective_masses}). In order to ascertain the suitability of this approach, we have used the conventional 6-band Luttinger-Kohn \textbf{k}$\cdot$\textbf{p} Hamiltonian \cite{Luttinger_PR_1955} in conjunction with the spherical approximation \cite{Andreani_PRB_1987} to estimate the LH DOS effective mass for comparison to our computed values of $m_{-}^{\ast}$. In the spherical approximation the LH effective mass is given by $m_{\scalebox{0.7}{\text{LH}}}^{\ast} = ( \gamma_{1} + 2 \overline{\gamma} )^{-1} m_{0}$, with $\overline{\gamma} = \frac{1}{5} ( 2 \gamma_{2} + 3 \gamma_{3} )$, and where $\gamma_{1}$, $\gamma_{2}$ and $\gamma_{3}$ are the VB Luttinger parameters. \cite{Luttinger_PR_1955} For InAs $\gamma_{1} = 20.0$, $\gamma_{2} = 8.5$ and $\gamma_{3} = 9.3$, \cite{Vurgaftman_JAP_2001} giving $m_{\scalebox{0.7}{\text{LH}}}^{\ast} = 0.0328 \, m_{0}$, while for InSb $\gamma_{1} = 34.8$, $\gamma_{2} = 15.5$ and $\gamma_{3} = 16.5$, \cite{Vurgaftman_JAP_2001} giving $m_{\scalebox{0.7}{\text{LH}}}^{\ast} = 0.0149 \, m_{0}$. We note that these values compare reasonably well with the values $m_{-}^{\ast} = 0.0274 \, m_{0}$ and $0.0140 \, m_{0}$ computed using Eq.~\eqref{eq:2_band effective_masses} for InAs and InSb respectively -- using the values of $E_{P}$ fit to $m_{+}^{\ast}$ above -- with Eq.~\eqref{eq:2_band effective_masses} correctly describing that the LH band edge effective mass is greater than that associated with the CB edge (i.e.~$m_{-}^{\ast} > m_{+}^{\ast}$).

\begin{figure}[t!]
	\includegraphics[width=0.94\columnwidth]{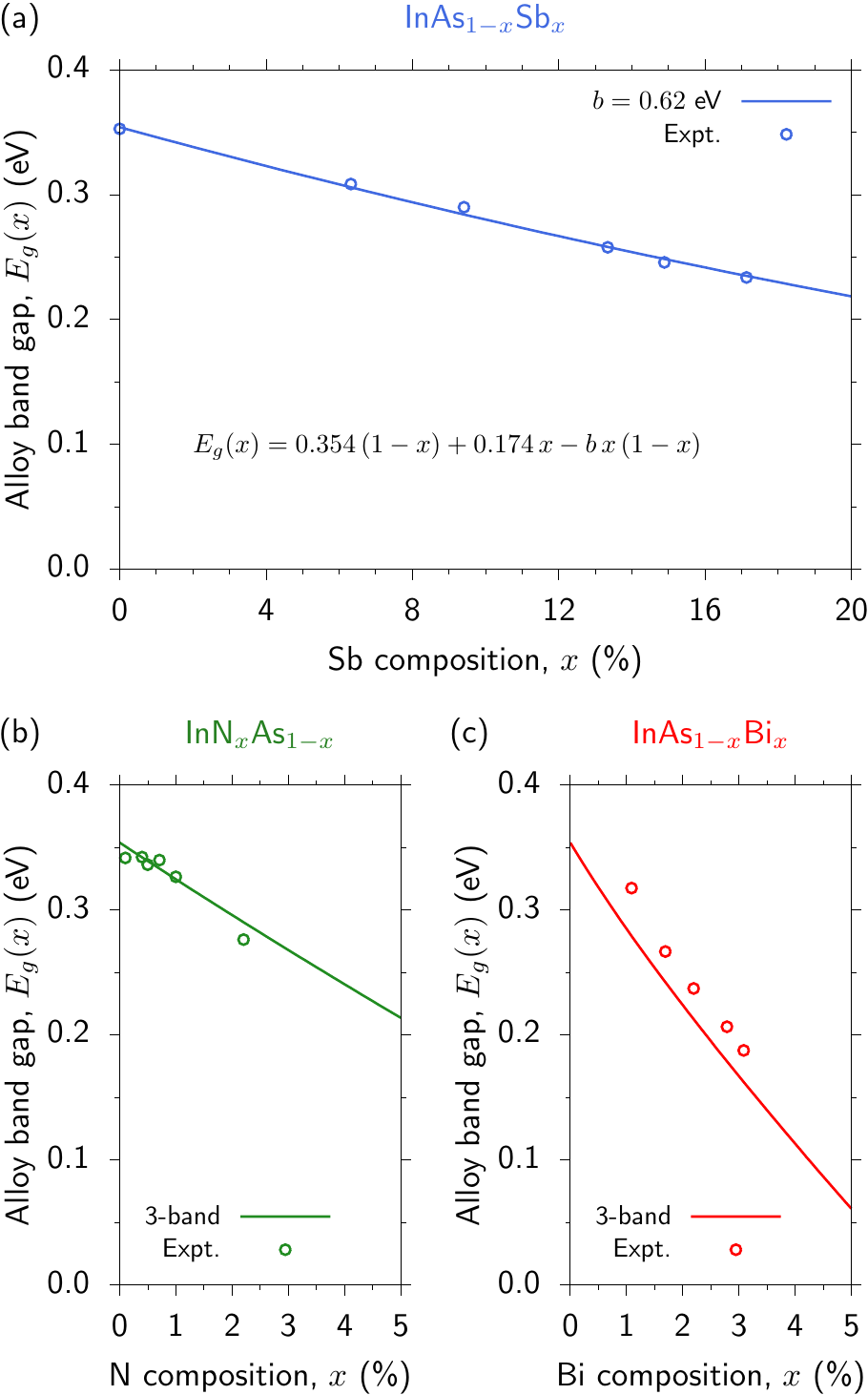}
	\caption{Calculated composition-dependent band gaps of (a) InAs$_{1-x}$Sb$_{x}$ (solid blue line), (b) InN$_{x}$As$_{1-x}$ (solid green line), and (c) InAs$_{1-x}$Bi$_{x}$ (solid red line). Open blue and green circles in (a) and (b) respectively denote the room temperature photoluminescence measurements of Refs.~\onlinecite{Murawski_PNS_2019} and~\onlinecite{Merrick_PRB_2007}. Open red circles in (c) denote the room temperature optical absorption measurements of Ref.~\onlinecite{Fang_JAP_1990}.}
	\label{fig:alloy_band_gaps}
\end{figure}

Finally, for InAs$_{1-x}$Sb$_{x}$ we interpolate the alloy band gap between that of InAs ($= 0.354$ eV) and InSb ($= 0.174$ eV) using a bowing parameter $b$ fit directly to experimental measurements. For the band gap range of interest in this study, which ranges from that of InAs down to $\approx 0.2$ eV, we have obtained $b = 0.62$ eV for InAs$_{1-x}$Sb$_{x}$ by fitting to the room temperature photoluminescence measurements of Ref.~\onlinecite{Murawski_PNS_2019} in the composition range $x \leq 20$\%. The result of this parametrisation is shown in Fig.~\ref{fig:alloy_band_gaps}(a), where we note that a composition-independent bowing parameter (solid blue line) is sufficient to quantitatively describe the measured alloy band gap (open blue circles) across the Sb composition range of interest.


To compute the complex band structure associated with Eq.~\eqref{eq:2_band_hamiltonian} we re-cast the characteristic equation as a quartic equation in $k_{z}$, which can be solved analytically as a quadratic equation in $k_{z}^{2}$ to yield the $E$- and $k_{\perp}$-dependent dispersion

\begin{equation}
    k_{z,\pm}^{2} ( E, k_{\perp} ) = \kappa_{\pm}^{2} (E) - k_{\perp}^{2} \, ,
    \label{eq:complex_wave_vector}
\end{equation}

\noindent
where the energy-dependent component $\kappa_{\pm}^{2} (E)$ is given by

\begin{equation}
    \frac{ \hbar^{2} \kappa_{\pm}^{2} }{ 2m_{0} } = E + \frac{ E_{P} - E_{g} }{ 2 } \pm \sqrt{ \left( E + \frac{ E_{P} - E_{g} }{ 2 } \right)^{2} - ( E - E_{g} ) \, E } \, .
    \label{eq:2_band_complex_bands}
\end{equation}

At fixed perpendicular wave vector $k_{\perp}$, the component $k_{z,\pm}$ of the wave vector parallel to the tunneling direction described by Eqs.~\eqref{eq:complex_wave_vector} and~\eqref{eq:2_band_complex_bands} is purely real for $E_{-} ( k_{\perp} ) \leq E$ and $E \geq E_{+} ( k_{\perp} )$, with $k_{z,-}$ reproducing the band dispersion of Eq.~\eqref{eq:2_band_real_bands} in this energy range. For $E_{-} ( k_{\perp} ) < E < E_{+} ( k_{\perp} )$ the band $k_{z,-}$ is purely imaginary. As such, $k_{z,-} ( E, k_{\perp} )$ describes the complex band linking the VB and CB at energies $E_{\pm} ( k_{\perp} )$. We note that $k_{z,+} ( E, k_{\perp} )$ is a spurious solution of Eq.~\eqref{eq:2_band_hamiltonian}, describing unphysical states which possess large real wave vectors and lie energetically within the band gap. \cite{Meney_PRB_1994}

Having described the calculation of the complex band structure for conventional InAs$_{1-x}$Sb$_{x}$ alloys, we now turn our attention to highly-mismatched InN$_{x}$As$_{1-x}$ and dilute bismide InAs$_{1-x}$Bi$_{x}$ alloys. When substituted in dilute concentrations in InAs to form dilute nitride InN$_{x}$As$_{1-x}$ isolated N atoms act as isovalent impurities, introducing a N-related localised impurity state which is resonant with the CB of the InAs host matrix semiconductor. \cite{Shan_PRL_1999,Veal_APL_2005,Kudrawiec_APL_2009} It is well established that the evolution of the main features of the InN$_{x}$As$_{1-x}$ band structure can then be described via a N composition-dependent BAC interaction between the N-related localised impurity state and the extended (Bloch) InAs CB edge state. This interaction pushes the alloy CB edge energy downwards in energy, accounting for the strong composition-dependent band gap bowing observed in experiment. \cite{Lindsay_SSC_1999,Lindsay_SSE_2003} Beginning from the 2-band \textbf{k}$\cdot$\textbf{p} Hamiltonian of Eq.~\eqref{eq:2_band_hamiltonian}, the resulting 3-band BAC Hamiltonian for InN$_{x}$As$_{1-x}$ at N composition $x$ is \cite{Reilly_SST_2002,Lindsay_SSE_2003}

\begin{equation}
	H = \left( \begin{array}{ccc}
    E_{\scalebox{0.6}{\text{N}}} & \beta \sqrt{x}                                                                               & 0                                                                                    \\
	\beta \sqrt{x}                 & E_{g} - \alpha \, x + \frac{ \hbar^{2} }{ 2 m_{0} } \left( k_{\perp}^{2} + k_{z}^{2} \right) & i P \, \sqrt{ k_{\perp}^{2} + k_{z}^{2} }                                            \\
 	0                              & -i P \, \sqrt{ k_{\perp}^{2} + k_{z}^{2} }                                                   & \gamma \, x + \frac{ \hbar^{2} }{ 2 m_{0} } \left( k_{\perp}^{2} + k_{z}^{2} \right) \\
     \end{array} \right) \, .
    \label{eq:3_band_hamiltonian_nitride}
\end{equation}

Here, $E_{\scalebox{0.7}{\text{N}}}$ is the energy of the N-related localised impurity state, $\beta$ is the BAC coupling parameter describing the strength $\beta \sqrt{x}$ of the N composition-dependent BAC interaction, \cite{Reilly_SST_2002,Lindsay_SSE_2003} and $\alpha$ and $\gamma$ are virtual crystal type terms respectively describing the N-induced energy shifts to the host matrix CB and VB edge energies. \cite{Reilly_SST_2002,Tomic_IEEEJSTQE_2003} The parameters $E_{g}$ and $P$ are assigned the values given above for the InAs host matrix semiconductor.

The characteristic equation associated with Eq.~\eqref{eq:3_band_hamiltonian_nitride} is cubic in the energy $E$, with numerical diagonalisation for real wave vectors $\textbf{k} = ( \textbf{k}_{\perp}, k_{z} )$ yielding the (real, spherical) band structure. However, since the N-related energy level is dispersionless, re-casting the characteristic equation in terms of $k_{z}$ again yields a quartic equation which can be factorised analytically. In this manner it is therefore possible to compute the complex band structure associated with Eq.~\eqref{eq:3_band_hamiltonian_nitride} analytically, yielding an expression for the complex band dispersion $k_{z,\pm}^{2} ( E, k_{\perp} ) = \kappa_{\pm}^{2} ( E ) - k_{\perp}^{2}$ (cf.~Eq.~\eqref{eq:complex_wave_vector}) where $\kappa_{\pm}^{2}$ is given by

\begin{widetext}
    \begin{equation}
        \frac{ \hbar^{2} \kappa_{\pm}^{2} }{ 2m_{0} } = E + \frac{ E_{P} - E_{g} + ( \alpha - \gamma ) x }{ 2 } + \frac{ \beta^{2} x }{ 2 ( E_{\scalebox{0.7}{\text{N}}} - E ) } \pm \sqrt{ \left( E + \frac{ E_{P} - E_{g} + ( \alpha - \gamma ) x }{ 2 } + \frac{ \beta^{2} x }{ 2 ( E_{\scalebox{0.7}{\text{N}}} - E ) } \right)^{2} - \left( E - E_{g} + \alpha x + \frac{ \beta^{2} x }{ E_{\scalebox{0.7}{\text{N}}} - E } \right) \left( E - \gamma x \right) } \, ,
        \label{eq:3_band_complex_bands_nitride}
    \end{equation}
\end{widetext}

\noindent
which reduces to Eq.~\eqref{eq:2_band_complex_bands} for $x = 0$.

The N-related parameters of Eq.~\eqref{eq:3_band_hamiltonian_nitride} have most often been estimated by diagonalising the BAC model at $k = 0$ in order to obtain analytical expressions for the composition-dependent band edge energies, with the parameters $E_{\scalebox{0.7}{\text{N}}}$ and $\beta$ then adjusted to fit to experimental band gap data. \cite{Shan_PRL_1999,Veal_APL_2005,Kudrawiec_APL_2009} However, this procedure is undesirable in general, because it typically relies on fitting two (or more) parameters to a single piece of experimental data (the band gap at a given N composition $x$), requiring assumptions to be applied regarding unconstrained parameters and thereby introducing systematic uncertainty. To avoid this, we have used atomistic alloy supercell electronic structure calculations to directly compute all N-related parameters -- $E_{\scalebox{0.7}{\text{N}}}$, $\beta$, $\alpha$ and $\gamma$ -- in Eq.~\eqref{eq:3_band_hamiltonian_nitride}, removing the requirement to employ post hoc fitting to experimental data. \cite{Lindsay_SSE_2003,Reilly_SST_2009,Broderick_nitride_chapter_2017} The resulting parameters are listed in Table~\ref{tab:parameters}. We note that the N localised state energy $E_{\scalebox{0.7}{\text{N}}}$ is given relative to the InAs host matrix VB edge -- which we maintain as our zero of energy throughout -- so that the N-related impurity level in InAs lies 1.032 eV ($= E_{\scalebox{0.7}{\text{N}}} - E_{g}$) above the CB edge in energy.

Using the N-related parameters of Table~\ref{tab:parameters}, we have calculated the composition-dependent InN$_{x}$As$_{1-x}$ alloy band gap $E_{g} (x)$ at $k = 0$, shown in Fig.~\ref{fig:alloy_band_gaps}(b) using a solid green line. Here, our calculations predict that N incorporation in InAs produces a band gap reduction of approximately 29 meV per \% N replacing As. Open green circles in Fig.~\ref{fig:alloy_band_gaps}(b) show the band gap determined via room temperature photoluminescence measurements. \cite{Merrick_PRB_2007} Our calculations are in good, quantitative agreement with these experimental data.


\begin{table}[t!]
	\caption{\label{tab:parameters} Parameters for the 3-band BAC Hamiltonians of InN$_{x}$As$_{1-x}$ and InAs$_{1-x}$Bi$_{x}$, Eqs.~\eqref{eq:3_band_hamiltonian_nitride} and~\eqref{eq:3_band_hamiltonian_bismide}, including the N- and Bi-related resonant impurity state energies $E_{\protect\scalebox{0.7}{\text{N}}}$ and $E_{\protect\scalebox{0.7}{\text{Bi}}}$, the BAC coupling strength $\beta$, and the CB and VB edge energy virtual crystal shifts $\alpha$ and $\gamma$. $E_{\protect\scalebox{0.7}{\text{N}}}$ and $E_{\protect\scalebox{0.7}{\text{Bi}}}$ are given relative to the zero of energy at the InAs VB maximum. All parameters are listed in units of eV.}
	\begin{ruledtabular}
		\begin{tabular}{ccc}
			Parameter                       & InN$_{x}$As$_{1-x}$ & InAs$_{1-x}$Bi$_{x}$ \\
			\hline
			$E_{\scalebox{0.7}{\text{N}}}$  & 1.386$^{a}$         &    -----       \\
			$E_{\scalebox{0.7}{\text{Bi}}}$ & -----               & $-$0.217$^{d}$ \\
			$\beta$                         & 1.212$^{a}$         &    0.920$^{d}$ \\
			$\alpha$                        & 1.550$^{b}$         &    2.603$^{d}$ \\
			$\gamma$                        & 0.000$^{c}$         &    1.027$^{d}$
		\end{tabular}
	\end{ruledtabular}
    \vspace{-0.2cm}
    \begin{flushleft}
    	    $^{a}$Ref.~\onlinecite{Reilly_SST_2009} \;
    	    $^{b}$Ref.~\onlinecite{Tomic_IEEEJSTQE_2003} \;
    	    $^{c}$Ref.~\onlinecite{Healy_IEEEJQE_2006} \;
    	    $^{d}$Ref.~\onlinecite{Chai_SST_2015} \;
    \end{flushleft}
\end{table}

Finally, we consider the complex band structure of dilute bismide InAs$_{1-x}$Bi$_{x}$ alloys. Here, the band structure mirrors that of InN$_{x}$As$_{1-x}$: while small, electronegative N atoms strongly perturb the CB structure in dilute nitride alloys, \cite{Shan_PRL_1999,Broderick_nitride_chapter_2017} in dilute bismide alloys large, electropositive Bi atoms strongly perturb the VB structure. \cite{Alberi_PRB_2007,Broderick_bismide_chapter_2017} When substituted in InAs, an isolated Bi atom acts as an isovalent impurity, introducing a Bi-related localised state which is resonant with the VB of the InAs host matrix semiconductor. \cite{Broderick_PSSB_2013} As in InN$_{x}$As$_{1-x}$, the evolution of the main features of the InAs$_{1-x}$Bi$_{x}$ band structure are well described by a BAC model, but in which the Bi-related localised impurity state undergoes a composition-dependent BAC interaction with the extended (Bloch) InAs VB edge state. This interaction pushes the alloy VB edge upwards in energy, accounting for the strong composition-dependent band gap bowing observed in experiment. \cite{Alberi_PRB_2007,Alberi_APL_2007,Usman_PRB_2011,Broderick_SST_2013} The associated 3-band BAC Hamiltonian for InAs$_{1-x}$Bi$_{x}$ at Bi composition $x$ is \cite{Broderick_SST_2013,Broderick_PSSB_2013}

\begin{equation}
	H = \left( \begin{array}{ccc}
    E_{g} - \alpha \, x + \frac{ \hbar^{2} }{ 2 m_{0} } \left( k_{\perp}^{2} + k_{z}^{2} \right) & i P \, \sqrt{ k_{\perp}^{2} + k_{z}^{2} }                                            & 0                               \\
	-i P \, \sqrt{ k_{\perp}^{2} + k_{z}^{2} }                                                   & \gamma \, x + \frac{ \hbar^{2} }{ 2 m_{0} } \left( k_{\perp}^{2} + k_{z}^{2} \right) & \beta \sqrt{x}                  \\
 	0                                                                                            & \beta \sqrt{x}                                                                       & E_{\scalebox{0.6}{\text{Bi}}} \\
    \end{array} \right) \, ,
    \label{eq:3_band_hamiltonian_bismide}
\end{equation}

\noindent
where $\beta$ now describes the strength $\beta \sqrt{x}$ of the composition dependent BAC interaction between the Bi-related localised impurity state (at energy $E_{\scalebox{0.7}{\text{Bi}}}$) and the InAs VB edge. \cite{Broderick_SST_2013} Again, $\alpha$ and $\gamma$ respectively describe Bi-induced virtual crystal shifts to the host matrix CB and VB edge energies, \cite{Broderick_SST_2013} and $E_{g}$ and $P$ are assigned values associated with the InAs host matrix semiconductor. As for InN$_{x}$As$_{1-x}$, the InAs$_{1-x}$Bi$_{x}$ complex band structure can be computed analytically by diagonalising the characteristic equation associated with Eq.~\eqref{eq:3_band_hamiltonian_bismide} analytically, again yielding the dispersion of Eq.~\eqref{eq:complex_wave_vector} with

\begin{widetext}
    \begin{equation}
        \frac{ \hbar^{2} \kappa_{\pm}^{2} }{ 2m_{0} } = E + \frac{ E_{P} - E_{g} + ( \alpha - \gamma ) x }{ 2 } + \frac{ \beta^{2} x }{ 2 ( E_{\scalebox{0.7}{\text{Bi}}} - E ) } \pm \sqrt{ \left( E + \frac{ E_{P} - E_{g} + ( \alpha - \gamma ) x }{ 2 } + \frac{ \beta^{2} x }{ 2 ( E_{\scalebox{0.7}{\text{Bi}}} - E ) } \right)^{2} - \left( E - E_{g} + \alpha x \right) \left( E - \gamma x + \frac{ \beta^{2} x }{ E_{\scalebox{0.7}{\text{Bi}}} - E } \right) } \, ,
        \label{eq:3_band_complex_bands_bismide}
    \end{equation}
\end{widetext}

\noindent
which also reduces to Eq.~\eqref{eq:2_band_complex_bands} for $x = 0$.

Again, we have computed the Bi-related parameters of Eq.~\eqref{eq:3_band_hamiltonian_bismide} directly via atomistic electronic structure calculations, \cite{Broderick_SST_2013,Broderick_PSSB_2013} giving the values listed for InAs$_{1-x}$Bi$_{x}$ in Table~\ref{tab:parameters}. Figure~\ref{fig:alloy_band_gaps}(c) shows the InAs$_{1-x}$Bi$_{x}$ alloy band gap $E_{g} (x)$ calculated via Eq.~\eqref{eq:3_band_hamiltonian_bismide} (solid red line), compared to that determined via room temperature optical absorption measurements (open red circles). \cite{Fang_JAP_1990} While our predicted band gap in Fig.~\ref{fig:alloy_band_gaps} appears to underestimate the measured band gap, we note that extrapolating the experimental data back to $x = 0$ produces an InAs band gap which is significantly in excess of that which would be expected at room temperature (cf.~open blue circle at $x = 0$ in Fig.~\ref{fig:alloy_band_gaps}(a)). \cite{Vurgaftman_JAP_2001} On this basis we expect that the Bi compositions in Ref.~\onlinecite{Fang_JAP_1990} are overestimated, resolution of which would bring our theoretical calculations into closer agreement with the measured band gaps. As such, we emphasise that our parametrisation of Eq.~\eqref{eq:3_band_hamiltonian_bismide} accurately describes the measured rate at which the alloy band gap reduces with increasing Bi composition $x$, with our calculated reduction of approximately 69 meV per \% Bi replacing As being in good, quantitative agreement with experiment.


\subsection{BTBT transmission coefficient and generation rate}
\label{sec:theory_btbt}

In the WKB approximation the transmission coefficient $T$ associated with direct BTBT at applied electric field $F = \vert \textbf{F} \vert$ is computed as \cite{Pan_JAP_2014,Esseni_SST_2017}

\begin{equation}
    T ( \textbf{k}_{\perp} ) = \frac{ \pi^{2} }{ 9 } \exp \left( - \frac{ 2 }{ eF } \int_{ E_{\scalebox{0.6}{\text{LH}}} ( \textbf{k}_{\perp} ) }^{ E_{\scalebox{0.6}{\text{CB}}} ( \textbf{k}_{\perp} ) } \left| \text{Im} \left\{ k_{z,-} ( E, \textbf{k}_{\perp} ) \right\} \right| \text{d} E \right)
    \label{eq:transmission_wkb}
\end{equation}

\noindent
where, for tunneling along [001] $\parallel$ $\textbf{F}$, $k_{z,-} ( E, \textbf{k}_{\perp} )$ is the complex band linking the VB and CB at perpendicular wave vector $\textbf{k}_{\perp}$ (cf.~Eq.~\eqref{eq:complex_wave_vector}), and the upper and lower limits on the integral are the energies of the CB and LH band at $\textbf{k}_{\perp}$. We note that $T$ does not depend explicitly on energy in the WKB approximation, unlike in more sophisticated quantum kinetic approaches. However, energy-averaged values of $T ( \textbf{k}_{\perp} )$ obtained from non-equilibrium Green's function calculations for bulk InAs have been demonstrated to be in good, quantitative agreement with Eq.~\eqref{eq:transmission_wkb}, validating the application of the WKB approximation. \cite{Pan_JAP_2014}

Given the BTBT transmission coefficient, the field-dependent BTBT generation rate (per unit volume) $G$ is computed via integration with respect to $\textbf{k}_{\perp}$ \cite{Esseni_SST_2017}

\begin{equation}
    G ( F ) = \frac{ e F }{ \pi \hbar } \int \frac{ \text{d} \textbf{k}_{\perp} }{ ( 2 \pi )^{2} } \, T ( \textbf{k}_{\perp} ) \, .
    \label{eq:generation_wkb}
\end{equation}

Recalling that the band structures admitted by Eqs.~\eqref{eq:2_band_hamiltonian},~\eqref{eq:3_band_hamiltonian_nitride} and~\eqref{eq:3_band_hamiltonian_bismide} are spherical -- i.e.~the band dispersion depends only on $k = \vert \textbf{k} \vert$ -- we can write $\text{d} \textbf{k}_{\perp} = 2 \pi k_{\perp} \text{d} k_{\perp}$, so that computation of $G$ via Eq.~\eqref{eq:generation_wkb} requires numerical quadrature in only one dimension. The generation rate $G$ is linked to the BTBT current density $J$ via $eGV = JF$, \cite{Luisier_IEEEEDL_2009,Esseni_SST_2017} where $V$ is the applied voltage generating the electric field $F$. Here, we focus solely on $G$ as an intrinsic material property, and simply note that defining a specific tunnel junction geometry and doping fixes the relationship between an applied voltage $V$ and the resulting (local) electric field $F$, enabling the $J$-$V$ relationship between the applied voltage and resulting BTBT current density for the junction to be computed straightforwardly once the generation rate $G$ is known.

Before proceeding, we emphasise the link between Eqs.~\eqref{eq:transmission_wkb} and~\eqref{eq:generation_wkb} and the widely employed Kane model for BTBT in a direct-gap semiconductor. \cite{Kane_JPCS_1959,Kane_JAP_1961} Given Kane's use of a semi-classical wave function phase in his analysis of BTBT, his approximate result for $T$ (and hence $G$) is formally equivalent to Eqs.~\eqref{eq:transmission_wkb} and~\eqref{eq:generation_wkb}. \cite{Pan_JAP_2014} Indeed, Kane's analytical result for $T$ can be straightforwardly obtained from Eq.~\eqref{eq:transmission_wkb} via two simple approximations: (i) that only band states close to $k_{\perp} = 0$ contribute appreciably to BTBT, and (ii) that the dispersion of the CB and LH bands is purely parabolic. \cite{Esseni_SST_2017} The first of these approximations enables the integral of Eq.~\eqref{eq:transmission_wkb} to be computed analytically, while the second ensures that the growth of the energy gap between the CB and LH bands at fixed $k_{\perp} \neq 0$ can be described straightforwardly in terms of a reduced effective mass $m_{r}^{\ast}$. The resulting expression for the transmission coefficient is \cite{Esseni_SST_2017}

\begin{equation}
    T ( k_{\perp} ) = \frac{ \pi^{2} }{ 9 } \exp \left( - \frac{ \pi \sqrt{ m_{r}^{\ast} } E_{g}^{3/2} }{ 2 e \hbar F } \right) \exp \left( - \frac{ \pi \hbar }{ 2 e F } \sqrt{ \frac{ E_{g} }{ m_{r}^{*} } } \, k_{\perp}^{2} \right) \, ,
    \label{eq:transmission_kane}
\end{equation}

\noindent
where $( m_{r}^{\ast} )^{-1} = ( m_{+}^{\ast} )^{-1} + ( m_{-}^{\ast} )^{-1}$ is the reduced zone-centre effective mass of the CB (``$+$'') and LH (``$-$'') bands, given in the context of the 2-band Hamiltonian of Eq.~\eqref{eq:2_band_hamiltonian} by $m_{r}^{\ast} = \frac{ E_{g} }{ 2 E_{P} } \, m_{0}$ (cf.~Eq.~\eqref{eq:2_band effective_masses}).

We note that Eq.~\eqref{eq:transmission_kane} is comprised of a product of two exponential terms, which respectively describe field-dependent zone-centre ($k_{\perp} = 0$) and off-zone-centre ($k_{\perp} \neq 0$) contributions to direct BTBT. The physical interpretation of these terms is useful to recapitulate, as it describes features which apply generally to direct BTBT in narrow-gap semiconductors, and will therefore aid in our interpretation of the impact of BAC on BTBT in Sec.~\ref{sec:results} below. The first of these terms, the zone-centre term, describes that $T$ is a strong function of the $k_{\perp} = 0$ band gap $E_{g}$, increasing exponentially with decreasing $E_{g}$ and with decreasing $m_{r}^{\ast}$. The second, off-zone-centre term describes that contributions to $T$ from band states at $k_{\perp} \neq 0$ diminish rapidly with increasing $k_{\perp}$, due to the quadratic increase with $k_{\perp}$ of the energy gap between the (assumed parabolic) CB and LH bands, as encapsulated by the zone-centre band gap $E_{g}$ and reduced effective mass $m_{r}^{\ast}$. We further note that the off-zone-centre term in Eq.~\eqref{eq:transmission_kane} is $\propto - k_{\perp}^{2} F^{-1}$, indicating that contributions to $T$ from band states at larger perpendicular wave vector become more important not only in the presence of small zone centre band gap $E_{g}$ and large reduced effective mass $m_{r}^{\ast}$, but also in the presence of high electric field strengths. As such, these two terms respectively encode the dependence of $T$, and hence $G$, on the band gap $E_{g}$ and the CB and VB edge DOS (encapsulated by $m_{r}^{\ast}$).

In Kane's model, the integral of Eq.~\eqref{eq:generation_wkb} can be computed analytically, yielding the BTBT generation rate \cite{Esseni_SST_2017}

\begin{equation}
    G ( F ) = \frac{ e^{2} F^{2} }{ 18 \pi \hbar^{2} } \, \sqrt{ \frac{ m_{r}^{\ast} }{ E_{g} } } \, \exp \left( - \frac{ \pi \sqrt{ m_{r}^{\ast} } E_{g}^{3/2} }{ 2 e \hbar F } \right) \, ,
    \label{eq:generation_kane}
\end{equation}

\noindent
which again displays strong dependence on the zone centre band gap $E_{g}$ and reduced effective mass $m_{r}^{\ast}$. In terms of nomenclature, we note that when we refer to the ``Kane model'' in the remainder of this paper we are referring to the analytical treatment of the BTBT transmission coefficient and generation rate summarised by Eqs.~\eqref{eq:transmission_kane} and~\eqref{eq:generation_kane}, and not to the 2-band Hamiltonian of Eq.~\eqref{eq:2_band_hamiltonian}.


\section{Results}
\label{sec:results}


\subsection{WKB approximation vs.~Kane model for InAs}
\label{sec:results_InAs}


\begin{figure*}[t!]
	\includegraphics[width=1.00\textwidth]{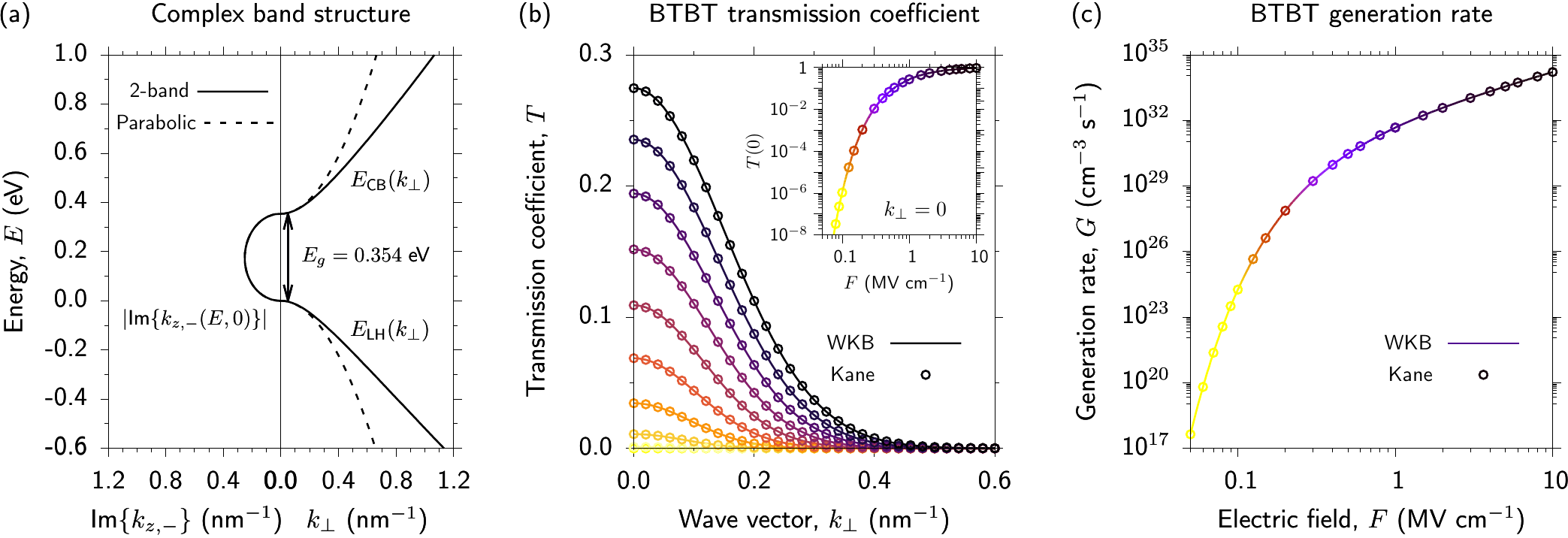}
	\caption{(a) Complex band structure of InAs calculated using the 2-band \textbf{k}$\cdot$\textbf{p} Hamiltonian of Eq.~\eqref{eq:2_band_hamiltonian} (solid lines), and the simplified parabolic bands employed in the derivation of the Kane model of direct BTBT (dashed lines). (b) BTBT transmission coefficient $T$ of InAs as a function of perpendicular wave vector $k_{\perp}$ calculated numerically using the 2-band complex dispersion of (a) in conjunction with the WKB approximation (solid lines), and using the analytical Kane model (open circles). Transmission coefficients are shown for applied electric fields $F$ ranging from 0.1 MV cm$^{-1}$ (yellow) to 1 MV cm$^{-1}$ (black) in steps of 0.1 MV cm$^{-1}$. Inset: zone-centre ($k_{\perp} = 0$) transmission coefficient $T$ as a function of applied electric field $F$, calculated numerically using the WKB approximation (solid line) and analytically using the Kane model (open circles). (c) BTBT generation rate $G$ of InAs as a function of applied electric field $F$, calculated numerically using the WKB approximation (solid line) and analytically using the Kane model (open circles).}
	\label{fig:btbt_InAs}
\end{figure*}

We begin by benchmarking our numerical calculations of the BTBT transmission coefficient $T$ and generation rate $G$ for InAs in the WKB approximation (Eqs.~\eqref{eq:transmission_wkb} and~\eqref{eq:generation_wkb}) via comparison to the analytical Kane model (Eqs.~\eqref{eq:transmission_kane} and~\eqref{eq:generation_kane}). The results of these calculations are summarised in Fig.~\ref{fig:btbt_InAs}. Beginning with the complex band structure in Fig.~\ref{fig:btbt_InAs}(a), the left- and right-hand panels respectively show the complex band dispersion $\text{Im} \lbrace k_{z,-} (E, 0) \rbrace$ at perpendicular wave vector $k_{\perp} = 0$, and the real band dispersion in the $k_{\perp}$ plane. The dashed black line in Fig.~\ref{fig:btbt_InAs}(a) depicts the simplified complex band structure employed in the derivation of the Kane model, while the solid black line shows the nonparabolic band structure described by the 2-band \textbf{k}$\cdot$\textbf{p} Hamiltonian of Eq.~\eqref{eq:2_band_hamiltonian} (cf.~Eqs.~\eqref{eq:complex_wave_vector} and~\eqref{eq:2_band_complex_bands}). The simplified band structure employed in the derivation of the Kane model is assumed to be parabolic in $k_{\perp}$, with the CB and LH bands having respective effective masses $m_{+}^{\ast}$ and $m_{-}^{\ast}$ (cf.~Eq.~\eqref{eq:2_band effective_masses}). We recall that the assumption of parabolic CB and LH dispersion allows the $k_{\perp}$-dependent band gap to be described solely in terms of the reduced effective mass $m_{r}^{\ast}$ (cf.~Eq.~\eqref{eq:transmission_kane}). For consistency between our numerical WKB and analytical Kane model calculations we therefore set $m_{r}^{\ast} = \frac{ E_{g} }{ 2 E_{P} } \, m_{0}$ in Eqs.~\eqref{eq:transmission_kane} and~\eqref{eq:generation_kane}, using the same values $E_{g} = 0.354$ eV and $E_{P} = 13.261$ eV for the band gap and Kane parameter of InAs that are employed in our 2-band \textbf{k}$\cdot$\textbf{p} calculation of the complex band structure. This ensures that both the numerical and analytical calculations employ band structures having equal band gap $E_{g}$ and band edge (zone-centre) effective masses $m_{\pm}^{\ast}$. We note that the complex band dispersion assumed in the derivation of the Kane model is the same as that obtained by diagonalising Eq.~\eqref{eq:2_band_hamiltonian} at $k_{\perp} = 0$ (cf.~Eq.~\eqref{eq:2_band_complex_bands}). The key difference between the \textbf{k}$\cdot$\textbf{p} band structure employed in our numerical WKB calculations and the simplified band structure employed in the derivation of the Kane model is then the presence of nonparabolicity in $k_{\perp}$ in the former.

Differences between the WKB and Kane model calculations of the BTBT transmission coefficient $T$ and generation rate $G$ are therefore expected to emerge only when band nonparabolicity is sufficiently pronounced to produce significant differences from the idealised parabolic band structure in the ranges of $k_{\perp}$ and $E$ at which states contribute to BTBT. Given the narrow band gap and associated low band edge effective masses in InAs we expect the impact of band nonparabolicity to be minimal, and hence that the results of our numerical WKB calculations should correspond closely to $T$ and $G$ computed using the Kane model. Figures~\ref{fig:btbt_InAs}(b) and~\ref{fig:btbt_InAs}(c) demonstrate that this is the case. Open circles in Fig.~\ref{fig:btbt_InAs}(b) show the analytical Kane model BTBT transmission coefficient $T$ as a function of $k_{\perp}$ computed using Eq.~\eqref{eq:transmission_kane}, while solid lines show the corresponding numerical WKB calculations using Eq.~\eqref{eq:transmission_wkb} (where Eq.~\eqref{eq:2_band_real_bands} sets the limits of integration, and Eqs.~\eqref{eq:complex_wave_vector} and~\eqref{eq:2_band_complex_bands} are used to evaluate the integrand). In Fig.~\ref{fig:btbt_InAs}(b) the colour denotes the strength $F$ of the applied electric field, varying from 0.1 MV cm$^{-1}$ (yellow) to 1.0 MV cm$^{-1}$ (black) in steps of 0.1 MV cm$^{-1}$. The inset to Fig.~\ref{fig:btbt_InAs}(b) shows the dependence of $T$ at $k_{\perp} = 0$ on applied electric field, demonstrating that the transmission coefficient for zone-centre BTBT rapidly approaches unity for high fields $F \gtrsim 1$ MV cm$^{-1}$. We note excellent quantitative agreement between our numerical WKB calculations of $T$ and the Kane model across the full ranges of $k_{\perp}$ and $F$ considered.

Finally, Fig.~\ref{fig:btbt_InAs}(c) shows the numerical WKB calculated BTBT generation rate $G$ as a function of $F$ (solid line), and the corresponding analytical Kane model calculation (open circles). The line colouring is as in Fig.~\ref{fig:btbt_InAs}(b). We again note excellent quantitative agreement between the two calculations. We also note that the calculated $G$ continues to grow exponentially as $F$ approaches 10 MV cm$^{-1}$. Recalling from the inset of Fig.~\ref{fig:btbt_InAs}(b) that the zone-centre transmission coefficient $T$ saturates at high electric fields, we emphasise that the continued exponential growth of $G$ at these high fields reflects that direct BTBT involving states at $k_{\perp} \neq 0$ constitutes an increasingly important contribution to $G$ as $F$ increases. As described in Sec.~\ref{sec:theory_btbt}, this is driven by the strong growth of $T$ away from $k_{\perp} = 0$ with increasing $F$ (cf.~Fig.~\ref{fig:btbt_InAs}(b) and Eq.~\eqref{eq:transmission_kane}). Overall, these benchmark calculations indicate that band nonparabolicity plays a negligible role in influencing direct BTBT in InAs, while also verifying the validity of our numerical WKB calculations as a platform to investigate BTBT in highly-mismatched alloys. Henceforth, all calculations of $T$ and $G$ are performed numerically in the WKB approximation, using Eqs.~\eqref{eq:transmission_wkb} and~\eqref{eq:generation_wkb} respectively.


\subsection{Fixed band gap: InAs$_{1-x}$Sb$_{x}$ vs.~InAs$_{1-x}$(N,Bi)$_{x}$ alloys}
\label{sec:results_fixed_band_gap}


\begin{figure*}[t!]
	\includegraphics[width=1.00\textwidth]{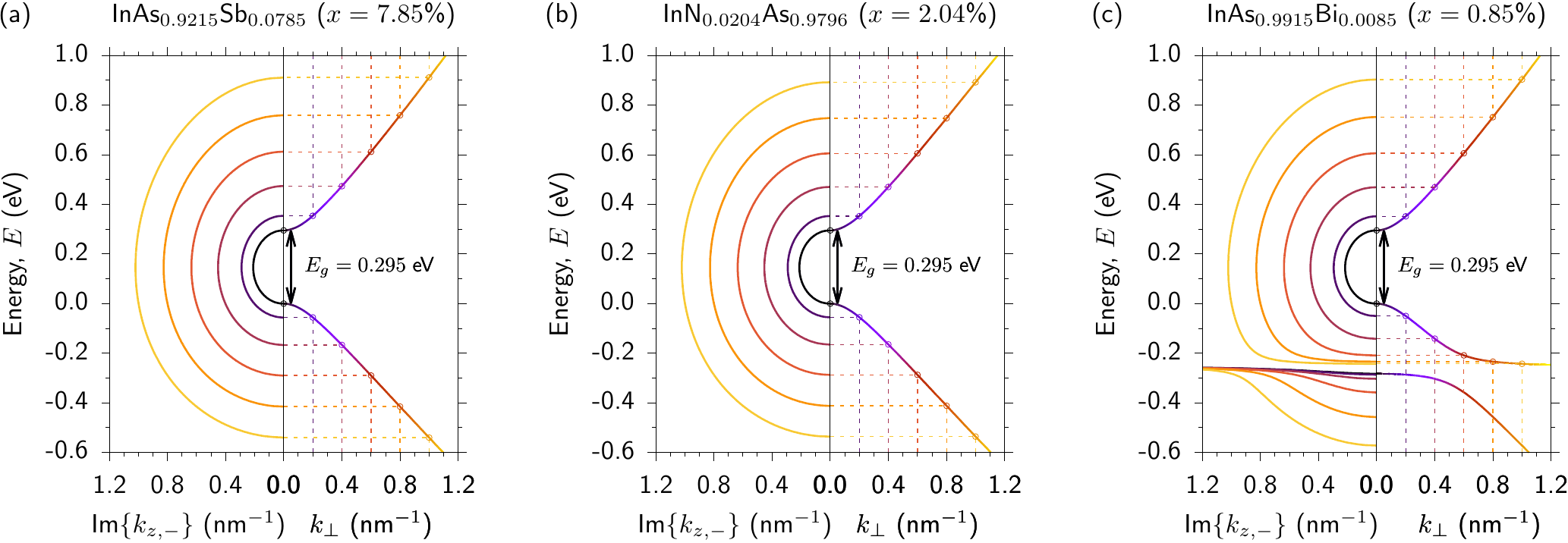}
	\caption{Calculated complex band structure of (a) InAs$_{1-x}$Sb$_{x}$ ($x = 7.85$\%), (b) InN$_{x}$As$_{1-x}$ ($x = 2.04$\%), and (c) InAs$_{1-x}$Bi$_{x}$ ($x = 0.85$\%) alloys having equal band gap $E_{g} = 0.295$ eV at $k_{\perp} = 0$. In each case, the left- (right-) hand panel shows the calculated imaginary (real) bands. The imaginary band $k_{z,-} ( E, k_{\perp} )$ is shown for perpendicular wave vectors $k_{\perp} =$ 0.0, 0.2, 0.4, 0.6, 0.8 and 1.0 nm$^{-1}$ using solid black, dark purple, purple, red, orange and yellow lines respectively. Dashed lines and open circles highlight that, for a given perpendicular wave vector $k_{\perp}$, the imaginary band $k_{z,-}$ links the CB and LH bands at energies $E_{\pm} ( k_{\perp} )$. Note that the features associated with the N-related localised impurity state in (b) are out of scale, since $E_{\protect\scalebox{0.7}{\text{N}}} = 1.386$ eV (cf.~Table~\ref{tab:parameters}).}
	\label{fig:4200_nm_band_structure}
\end{figure*}

As described above, while the complex band dispersion close to $k_{\perp} = 0$ dominates $G$ at low field, the evolution of $G$ for high electric fields $F \gtrsim 1$ MV cm$^{-1}$ is primarily governed by an interplay between the real and complex band dispersion away from $k_{\perp} = 0$. It is the strong modification of the real and complex band dispersion in highly-mismatched alloys due to BAC interactions that is expected to alter BTBT compared to that in a conventional alloy having the same band gap. To quantify these effects we begin by analysing $T$ and $G$ in highly-mismatched and conventional alloys at fixed band gap. Specifically, we consider highly-mismatched InN$_{x}$As$_{1-x}$ and InAs$_{1-x}$Bi$_{x}$ alloys, which we compare to InAs$_{1-x}$Sb$_{x}$, where in all three cases the alloy composition is chosen to ensure equal band gap $E_{g} = 0.295$ eV at $k_{\perp} = 0$. This exemplar band gap corresponds to a wavelength of 4.2 $\mu$m, chosen since it coincides with a strong absorption band of CO$_{2}$ and hence represents an important wavelength for mid-infrared sensing applications. Here, since we consider alloys having equal band gap, changes in $T$ and $G$ result solely from differences in the real and complex band dispersion, allowing to directly interrogate and quantify the impact of BAC on the BTBT transmission coefficient $T$ and generation rate $G$.

Figure~\ref{fig:4200_nm_band_structure} shows the calculated complex band structures of (a) InAs$_{1-x}$Sb$_{x}$ ($x = 7.85$\%), (b) InN$_{x}$As$_{1-x}$ ($x = 2.04$\%), and (c) InAs$_{1-x}$Bi$_{x}$ ($x = 0.85$\%) alloys having $E_{g} = 0.295$ eV. As in Fig.~\ref{fig:btbt_InAs}(a) the left- and right-hand panels respectively show the complex and real band dispersion, while the line colour denotes the magnitude $k_{\perp}$ of the perpendicular wave vector. Solid lines show the computed band structure calculated using Eq.~\eqref{eq:complex_wave_vector} in conjunction with (a) Eq.~\eqref{eq:2_band_complex_bands}, (b) Eq.~\eqref{eq:3_band_complex_bands_nitride}, and (c) Eq.~\eqref{eq:3_band_complex_bands_bismide}. The complex band dispersion $\text{Im} \lbrace k_{z,-} ( E, k_{\perp} ) \rbrace$ is shown in each case for perpendicular wave vectors $k_{\perp} = 0$ (black) to 1.0 nm$^{-1}$ (yellow) in steps of 0.2 nm$^{-1}$. Dashed lines and open circles indicate schematically the intersection between the real and complex bands at the CB and LH energies $E_{\pm} ( k_{\perp} )$ -- i.e.~the limits of integration in Eq.~\eqref{eq:transmission_wkb} -- obtained for InAs$_{1-x}$Sb$_{x}$ via Eq.~\eqref{eq:2_band_real_bands}, and for InN$_{x}$As$_{1-x}$ and InAs$_{1-x}$Bi$_{x}$ by diagonalising Eqs.~\eqref{eq:3_band_hamiltonian_nitride} and~\eqref{eq:3_band_hamiltonian_bismide} respectively. This highlights the growth of the area bounded by the complex band $\text{Im} \lbrace k_{z,-} ( E, k_{\perp} ) \rbrace$ linking the CB and VB with increasing $k_{\perp}$, which plays a critical role in determining the magnitude of $T$ at high electric field.

Beginning with dilute nitride InN$_{x}$As$_{1-x}$, we note that the calculated complex band structure of Fig.~\ref{fig:4200_nm_band_structure}(b) is minimally changed compared to that of InAs$_{1-x}$Sb$_{x}$ in Fig.~\ref{fig:4200_nm_band_structure}(a). Recalling the parameters of Table~\ref{tab:parameters}, we note that this is a consequence of InN$_{x}$As$_{1-x}$ representing a weakly perturbed highly-mismatched alloy: the N-related localised impurity state lies 1.032 eV above the InAs CB edge in energy, a large energy difference which minimises the impact of N-related BAC on the band structure close in energy to the CB edge in InN$_{x}$As$_{1-x}$. This can be seen in Eq.~\eqref{eq:3_band_complex_bands_nitride}, where the N composition-dependent perturbing terms at energy $E$ are inversely proportional to $E_{\scalebox{0.6}{\text{N}}} - E$. We compute a minimal increase in CB edge effective mass for InN$_{x}$As$_{1-x}$, leading to a slight reduction in the band gap at fixed $k_{\perp}$ compared to InAs$_{1-x}$Sb$_{x}$. However, considering the complex band structure of dilute bismide InAs$_{1-x}$Bi$_{x}$ in Fig.~\ref{fig:4200_nm_band_structure}(c), we note extremely strong modification compared to InAs$_{1-x}$Sb$_{x}$. This is a consequence of the close proximity in energy of the Bi-related localised impurity state in InAs to the VB edge (cf.~Table~\ref{tab:parameters}), resulting in a strongly perturbed VB structure (cf.~Eq.~\eqref{eq:3_band_complex_bands_bismide}). At small $k_{\perp}$ the BAC-induced increase in VB edge effective mass increases the area bounded by $\text{Im} \lbrace k_{z,-} ( E, k_{\perp} ) \rbrace$. At large $k_{\perp}$ the growth of this area is strongly limited in the VB by BAC: the dispersion of the complex band flattens close in energy to the Bi impurity state at $E_{\scalebox{0.6}{\text{Bi}}} = -0.217$ eV, the energy at which the real band dispersion admitted by Eq.~\eqref{eq:3_band_hamiltonian_bismide} possesses a singularity in the VB DOS.


\begin{figure*}[t!]
	\includegraphics[width=1.00\textwidth]{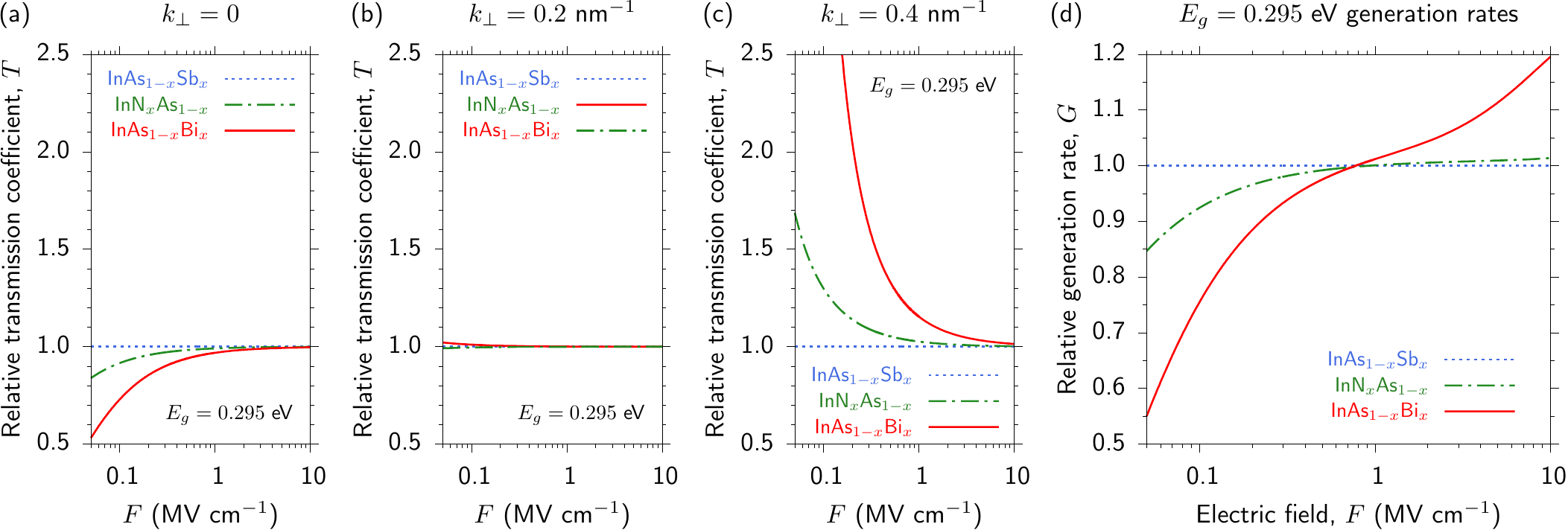}
	\caption{Calculated relative BTBT transmission coefficient $T$ at (a) $k_{\perp} = 0$, (b) $k_{\perp} = 0.2$ nm$^{-1}$, and (c) $k_{\perp} = 0.4$ nm$^{-1}$ for InAs$_{1-x}$Sb$_{x}$ ($x = 7.85$\%; dashed blue lines), InN$_{x}$As$_{1-x}$ ($x = 2.04$\%, dash-dotted green lines), and InAs$_{1-x}$Bi$_{x}$ ($x = 0.85$\%; solid red lines) alloys having equal band gap $E_{g} = 0.295$ eV at $k_{\perp} = 0$. (d) Calculated relative BTBT generation rate $G$ for InAs$_{1-x}$Sb$_{x}$ ($x = 7.85$\%; dotted blue lines), InN$_{x}$As$_{1-x}$ ($x = 2.04$\%, dash-dotted green lines), and InAs$_{1-x}$Bi$_{x}$ ($x = 0.85$\%; solid red lines) alloys. Relative values of $T$ and $G$ in (a) -- (d) are computed at each applied electric field $F$ by dividing the calculated value of $T$ or $G$ by that calculated for InAs$_{1-x}$Sb$_{x}$.}
	\label{fig:4200_nm_transmission_generation}
\end{figure*}

Based on the complex band structures of Fig.~\ref{fig:4200_nm_band_structure} we therefore expect that BAC in InN$_{x}$As$_{1-x}$ (InAs$_{1-x}$Bi$_{x}$) will, compared to InAs$_{1-x}$Sb$_{x}$, result in (i) reduced $T$ and hence $G$ at low $F$, where the increase in CB (VB) edge effective mass will act to increase the area bounded by $\text{Im} \lbrace k_{z,-} ( E, k_{\perp} ) \rbrace$ close to $k_{\perp} = 0$, and (ii) increased $T$ and hence $G$ at high $F$, where BAC will act to limit the growth of the area bounded by $\text{Im} \lbrace k_{z,-} ( E, k_{\perp} ) \rbrace$ at larger $k_{\perp}$. The latter effect is expected to be minimal in InN$_{x}$As$_{1-x}$, since the singularity in the CB DOS at energy $E_{\scalebox{0.6}{\text{N}}}$ occurs at large values of $k_{\perp}$ -- out of scale in Fig.~\ref{fig:4200_nm_band_structure}(b) -- at which the band gap is too large for states to contribute appreciably to BTBT. These expected trends are confirmed in Fig.~\ref{fig:4200_nm_transmission_generation}, which shows $T$ as a function of $F$ calculated at (a) $k_{\perp} = 0$, (b) $k_{\perp} = 0.2$ nm$^{-1}$, and (c) $k_{\perp} = 0.4$ nm$^{-1}$ for InAs$_{1-x}$Sb$_{x}$ (dashed blue lines), InN$_{x}$As$_{1-x}$ (dash-dotted green lines), and InAs$_{1-x}$Sb$_{x}$ (solid red lines). In each case $T$ is shown relative to that calculated for InAs$_{1-x}$Sb$_{x}$ at a given electric field $F$, allowing to directly quantify the impact of the aforementioned modifications of the complex band structure on BTBT at fixed band gap.

At $k_{\perp} = 0$, in Fig.~\ref{fig:4200_nm_transmission_generation}(a), we find that $T$ in both InN$_{x}$As$_{1-x}$ and InAs$_{1-x}$Bi$_{x}$ is reduced compared to that in InAs$_{1-x}$Sb$_{x}$. For example, at $F = 0.1$ MV cm$^{-1}$ $T$ in InN$_{x}$As$_{1-x}$ (InAs$_{1-x}$Bi$_{x}$) is reduced to 91.6\% (72.9\%) of that in InAs$_{1-x}$Sb$_{x}$, reflecting the stronger increase in band edge effective mass due to Bi incorporation, while at $F = 1$ MV cm$^{-1}$ $T$ in InN$_{x}$As$_{1-x}$ (InAs$_{1-x}$Bi$_{x}$) is 99.1\% (96.9\%) of that in InAs$_{1-x}$Sb$_{x}$, reflecting the saturation of $T$ at high field (cf.~Fig.~\ref{fig:btbt_InAs}(b)). At $k_{\perp} = 0.2$ nm$^{-1}$, in Fig.~\ref{fig:4200_nm_transmission_generation}(b), we note minimal differences between the values of $T$ calculated for the three alloys. We identify this value of the perpendicular wave vector as that corresponding to the crossover between ``small'' and ``large'' $k_{\perp}$, demarcating the crossover between the portions of the complex band structure that govern the behaviour of $G$ at ``low'' and ``high'' applied electric fields. Finally, at $k_{\perp} = 0.4$ nm$^{-1}$, in Fig.~\ref{fig:4200_nm_transmission_generation}(c), we find that $T$ in both InN$_{x}$As$_{1-x}$ and InAs$_{1-x}$Bi$_{x}$ is increased compared to that in InAs$_{1-x}$Sb$_{x}$, reflecting that BAC limits the growth of the area bounded by $\text{Im} \lbrace k_{z,-} ( E, k_{\perp} ) \rbrace$ at large $k_{\perp}$. Again, this effect is most pronounced for InAs$_{1-x}$Bi$_{x}$ due to the strength of the BAC-induced perturbation of the VB structure due to Bi incorporation (cf.~Fig.~\ref{fig:4200_nm_band_structure}(c)).

Finally, we consider the resulting electric field-dependent BTBT generation rate $G$ in Fig.~\ref{fig:4200_nm_transmission_generation}(d), shown for InAs$_{1-x}$Sb$_{x}$, InN$_{x}$As$_{1-x}$ and InAs$_{1-x}$Bi$_{x}$ using dashed blue, dash-dotted green and solid red lines, respectively. Here, $G$ is again shown relative to InAs$_{1-x}$Sb$_{x}$, by dividing $G$ calculated at each value of $F$ by that calculated for InAs$_{1-x}$Sb$_{x}$. Here, the calculated trends in the relative BTBT transmission coefficients of Figs.~\ref{fig:4200_nm_transmission_generation}(a) --~\ref{fig:4200_nm_transmission_generation}(c) are manifested. For InN$_{x}$As$_{1-x}$ we observe a reduction in $G$ at low field, with the calculated generation rate at $F = 0.1$ MV cm$^{-1}$ being 92.5\% of that in InAs$_{1-x}$Sb$_{x}$. By $F = 1$ MV cm$^{-1}$ the generation rate in InN$_{x}$As$_{1-x}$ is found to be within 0.1\% of that in InAs$_{1-x}$Sb$_{x}$. At higher fields $G$ in InN$_{x}$As$_{1-x}$ is calculated to exceed that in InAs$_{1-x}$Sb$_{x}$, but only minimally: at $F = 10$ MV cm$^{-1}$ $G$ in InN$_{x}$As$_{1-x}$ exceeds that in InAs$_{1-x}$Sb$_{x}$ by only 1.4\%. We recall from our discussion of the InN$_{x}$As$_{1-x}$ complex band structure that this behaviour is expected due to the large separation in energy between the InAs CB edge and the N-related impurity state, which results in minimal modification of the band dispersion in the ranges of $E$ and $k_{\perp}$ that contribute to BTBT. Turning our attention to InAs$_{1-x}$Bi$_{x}$ we find, in line with the calculated transmission coefficients, that the impact on $G$ due to Bi incorporation is enhanced compared to that associated with N incorporation. At low electric field $G$ is suppressed, being only 75.5\% of that in InAs$_{1-x}$Sb$_{x}$ at $F = 0.1$ MV cm$^{-1}$. We again find that $G$ is approximately equal to that in InAs$_{1-x}$Sb$_{x}$ for $F \approx 1$ MV cm$^{-1}$, beyond which it is strongly enhanced in the ``high-field'' regime, with $G$ at $F = 10$ MV cm$^{-1}$ exceeding that in InAs$_{1-x}$Sb$_{x}$ by 19.6\%.

This analysis highlights that incorporating N or Bi has the potential to significantly impact BTBT at fixed band gap compared to a conventional semiconductor alloy. Generally, BAC acts to suppress (enhance) the BTBT generation rate $G$ at low (high) electric field. Our analysis of the complex band structure identifies that this behaviour is governed by a field-dependent competition between various aspects of the band dispersion. At low fields $F \lesssim 1$ MV cm$^{-1}$ the BAC-induced increase in band edge effective mass increases the area bounded by $\text{Im} \lbrace k_{z,-} ( E, k_{\perp} ) \rbrace$ at small $k_{\perp}$, acting to suppress $T$ and hence $G$. At high fields $F \gtrsim 1$ MV cm$^{-1}$ the BAC-induced increase in DOS limits the growth of the band gap as $k_{\perp}$ increases, simultaneously limiting the growth of the area bounded by $\text{Im} \lbrace k_{z,-} ( E, k_{\perp} ) \rbrace$ at large $k_{\perp}$, acting to enhance $T$ and hence $G$. These effects are primarily governed by the strength of the impact of N- (Bi-) related BAC on the complex band structure close in energy to the CB (VB) edge, and are most pronounced when the associated N- (Bi-) related localised impurity state lies close in energy to the CB (VB) edge of the host matrix semiconductor.


\subsection{Variable band gap: InAs$_{1-x}$Sb$_{x}$ vs.~InAs$_{1-x}$(N,Bi)$_{x}$ alloys}
\label{sec:results_variable_band_gap}

\begin{figure}[t!]
	\includegraphics[width=0.92\columnwidth]{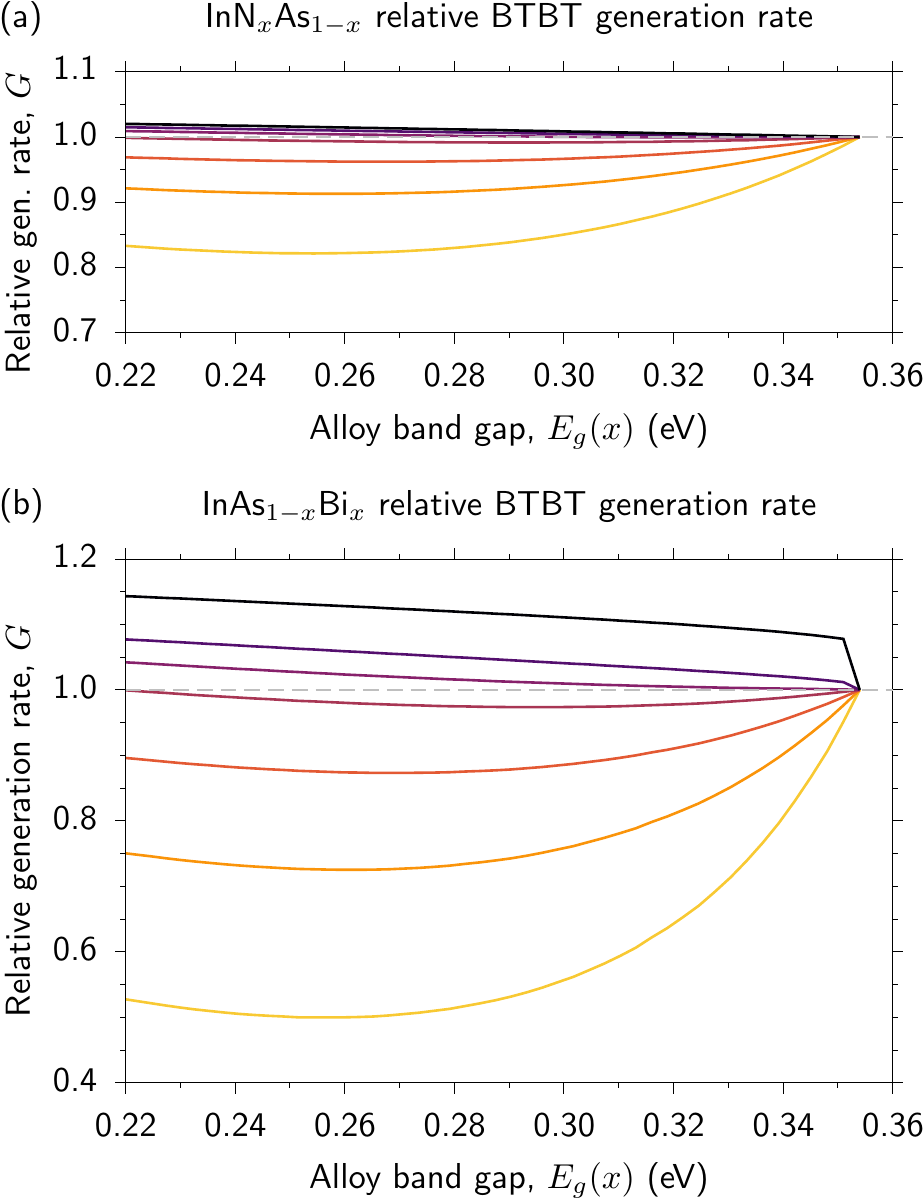}
	\caption{Calculated relative BTBT generation rate $G$ as a function of the alloy band gap $E_{g} (x)$ for (a) InN$_{x}$As$_{1-x}$, and (b) InAs$_{1-x}$Bi$_{x}$. Results are shown for applied electric fields $F =$ 50 kV cm$^{-1}$ (yellow), 0.1 MV cm$^{-1}$ (orange), 0.2 MV cm$^{-1}$ (red), 0.5 MV cm$^{-1}$ (magenta), 1 MV cm$^{-1}$ (purple), 2 MV cm$^{-1}$ (dark purple), and 5 MV cm$^{-1}$ (black). Relative values of $G$ are computed at each alloy band gap $E_{g} (x)$ by dividing the value calculated for InN$_{x}$As$_{1-x}$ or InAs$_{1-x}$Bi$_{x}$ by that calculated for an InAs$_{1-x}$Sb$_{x}$ alloy having the same band gap.}
	\label{fig:generation_rate_vs_band_gap}
\end{figure}

Having investigated the impact of BAC on BTBT at a single fixed band gap, we now turn our attention to trends as a function of the alloy band gap. The results of these calculations are summarised in Fig.~\ref{fig:generation_rate_vs_band_gap}, for (a) dilute nitride InN$_{x}$As$_{1-x}$, and (b) dilute bismide InAs$_{1-x}$Bi$_{x}$. In each case the BTBT generation rate $G$ is shown relative to that in InAs$_{1-x}$Sb$_{x}$, obtained by dividing $G$ calculated for InN$_{x}$As$_{1-x}$ and InAs$_{1-x}$Bi$_{x}$ by that calculated for an InAs$_{1-x}$Sb$_{x}$ alloy having the same band gap. Coloured solid lines in Fig.~\ref{fig:generation_rate_vs_band_gap} show the variation of the relative $G$ with the $k_{\perp} = 0$ alloy band gap $E_{g}$ for electric fields varying from 50 kV cm$^{-1}$ (yellow) to 5 MV cm$^{-1}$ (black). The dashed grey line in each case highlights the relative value $G = 1$ for the conventional (reference) InAs$_{1-x}$Sb$_{x}$ alloy.

In Fig.~\ref{fig:generation_rate_vs_band_gap} we see more broadly the trends identified via our fixed band gap calculations above. For low fields $F \lesssim 1$ MV cm$^{-1}$ (solid yellow, orange, red and magenta lines) the BTBT generation rate in InN$_{x}$As$_{1-x}$ and InAs$_{1-x}$Bi$_{x}$ is suppressed compared to that in InAs$_{1-x}$Sb$_{x}$. For the lowest electric field $F = 50$ kV cm$^{-1}$ considered, we calculate that the relative $G$ is minimised in InN$_{x}$As$_{1-x}$ (InAs$_{1-x}$Bi$_{x}$) for an alloy band gap of 0.254 eV (0.257 eV) -- corresponding to a N (Bi) composition $x = 3.50$\% (1.45\%) -- where it attains a value equal to 82.2\% (50.0\%) of that in InAs$_{1-x}$Sb$_{x}$. Again, we note that the reduction of $G$ at low field is significantly more pronounced in InAs$_{1-x}$Bi$_{x}$ than in InN$_{x}$As$_{1-x}$. The observed dependence of the calculated relative $G$ values on alloy band gap can be understood in terms of the competing impacts of the BAC-induced band gap narrowing and increase in CB (VB) edge effective mass in InN$_{x}$As$_{1-x}$ (InAs$_{1-x}$Bi$_{x}$). In conventional InAs$_{1-x}$Sb$_{x}$ the decrease in $E_{g}$ with increasing Sb composition is associated with a decrease in reduced effective mass $m_{r}^{\ast}$ ($\propto E_{g}$), both of which act to increase $G$ at fixed electric field (cf.~Eq.~\eqref{eq:generation_kane}) by reducing the area bounded by $\text{Im} \lbrace k_{z,-} ( E, k_{\perp} ) \rbrace$. However, in highly-mismatched dilute nitride (bismide) alloys the reduction in $E_{g}$ is generally associated with an increase in CB (VB) edge effective mass, with the former and latter respectively acting to increase and decrease $G$. Since at low $F$ contributions to $G$ are dominated by states close to $k_{\perp} = 0$, the BAC-induced increase in band edge effective mass plays the decisive role and leads to an initial reduction in the relative $G$ with decreasing $E_{g}$ in InN$_{x}$As$_{1-x}$ and InAs$_{1-x}$Bi$_{x}$. As $E_{g}$ decreases further, the band gap narrowing acts to increase $G$ with sufficient strength that it begins to counteract the tendency of the increased band edge effective mass to reduce $G$. This leads to a minimum in the calculated band gap dependence of the relative $G$ BTBT generation rate at low field. However, we note overall that the calculated low-field values of $G$ for InN$_{x}$As$_{1-x}$ and InAs$_{1-x}$Bi$_{x}$ in this $E_{g} \lesssim 0.25$ eV range remain lower than those calculated for InAs$_{1-x}$Sb$_{x}$ having equal band gap, confirming that the tendency of BAC to decrease low-field BTBT current holds across the full range of band gaps considered in our analysis.


For high fields $F \gtrsim 1$ MV cm$^{-1}$ (solid dark purple and black lines), we see that the BTBT generation rate in InN$_{x}$As$_{1-x}$ and InAs$_{1-x}$Bi$_{x}$ uniformly exceeds that in InAs$_{1-x}$Sb$_{x}$. As described above, the amount by which $G$ in InN$_{x}$As$_{1-x}$ exceeds that in InAs$_{1-x}$Sb$_{x}$ is limited by the weak impact of N-related BAC on the CB structure in the ranges of $E$ and $k_{\perp}$ that contribute to BTBT. Our calculations in Fig.~\ref{fig:generation_rate_vs_band_gap}(a) reveal that this behaviour is weakly dependent on the band gap, so that high-field BTBT currents in InN$_{x}$As$_{1-x}$ are expected to be close to those in InAs$_{1-x}$Sb$_{x}$ across the range of dilute N compositions relevant to device applications. Turning to InAs$_{1-x}$Bi$_{x}$ in Fig.~\ref{fig:generation_rate_vs_band_gap}(b), we note here an unusual feature of our calculations: at high fields $F \gtrsim 1$ MV cm$^{-1}$ our calculations suggest an abrupt increase in $G$ due to Bi incorporation. We note that this ``jump'' in $G$ is a feature of the model employed rather than a true physical effect: it is a consequence of a well-known unphysical aspect of the BAC model. Specifically, the LH band dispersion described by the BAC model of Eq.~\eqref{eq:3_band_hamiltonian_bismide} approaches the Bi localised impurity state energy $E_{\scalebox{0.6}{\text{Bi}}}$ asymptotically with increasing $k_{\perp}$ (cf.~Fig.~\ref{fig:4200_nm_band_structure}(c)). This produces a singularity in the DOS at energy $E_{\scalebox{0.6}{\text{Bi}}}$, below which energy there exists a gap in the DOS. In the context of the present calculations, the origin of this ``jump'' in $G$ is the asymptotic dispersion of the LH band: the flattening of this band with increasing $k_{\perp}$, as its energy approaches $E_{\scalebox{0.6}{\text{Bi}}}$, produces a range of states at large $k_{\perp}$ that can contribute to BTBT with increasing $F$. The magnitude of this ``jump'' in $G$ is then a consequence of the cut-off value of $k_{\perp}$ employed in the numerical evaluation of Eq.~\eqref{eq:generation_wkb}.

These unphysical aspects of the BAC model have been understood via theoretical analysis of GaN$_{x}$As$_{1-x}$ and related dilute nitride alloys. The established approach to circumvent these issues has been to treat the BAC interaction via a Green's function solution of the Anderson impurity model. \cite{Wu_PRB_2002,Vaughan_PRB_2007} This approach assigns an energy broadening to the localised impurity state, removing the unphysical singularity and gap in the DOS predicted by the BAC model. Generally, this broadening has been treated phenomenologically, and assigned a constant value (which can be estimated using perturbation theory \cite{Vaughan_PRB_2007,Seifikar_PSSB_2011}). More rigorous analysis of the Anderson impurity model has demonstrated that this localised state broadening is energy dependent, requiring a self-consistent computation of the Green's function. \cite{Jauho_PRB_1983} In a simple BAC model treating localised states having a single energy -- such as the models employed in this work -- this self-consistent Green's function approach removes the singularity in the DOS at the impurity energy, but the aforementioned gap in the DOS remains. \cite{Seifikar_PSSB_2011} Rigorous self-consistent calculations for GaN$_{x}$As$_{1-x}$ demonstrate that this gap in the DOS is removed only when a full distribution of N-related localised (cluster) states is treated explicitly, producing a DOS which is in agreement with that obtained from large-scale atomistic supercell calculations. \cite{Seifikar_JPCM_2014} To integrate such an approach into rigorous calculations of the BTBT generation rate is beyond the scope of this work. Therefore, while there is an uncertainty in our prediction of the degree to which $G$ in InAs$_{1-x}$Bi$_{x}$ exceeds that in InAs$_{1-x}$Sb$_{x}$, we emphasise that an increase is to be expected at higher fields. This is because, even in the absence of an unphysical singularity in the DOS at the impurity state energy, Green's function solutions of the Anderson impurity model uniformly predict that the presence of BAC nonetheless produces a significant enhancement of the DOS close in energy to the impurity state. So, while our simple BAC model likely overestimates the magnitude of this increase in DOS, the fact that the DOS in InAs$_{1-x}$Bi$_{x}$ is expected to significantly exceed that of the InAs host matrix semiconductor close in energy to $E_{\scalebox{0.6}{\text{Bi}}}$ indicates that our primary conclusions regarding the nature of high-field BTBT should hold.

Overall, our calculations reveal the existence of distinct low-field ($F \lesssim 1$ MV cm$^{-1}$) and high-field ($F \gtrsim 1$ MV cm$^{-1}$) regimes for BTBT in ideal highly-mismatched alloys. In the low-field regime, where tunneling is dominated by states close to $k_{\perp} = 0$, the BAC-induced increase in band edge effective mass dominates and acts to reduce BTBT current compared to that in a conventional semiconductor having the same band gap. In the high-field regime, where tunneling is dominated by states at large $k_{\perp}$, the BAC-induced modification of the real DOS and complex band dispersion dominates and acts to increase the BTBT current. Additionally, our analysis highlights that careful theoretical treatment of BTBT is required in cases where localised impurity states lie close in energy to the CB or VB edge of the host matrix semiconductor. In this case, the unphysical prediction by the BAC model of a singular DOS at the impurity energy precludes fully quantitative prediction of the BTBT current when states close in energy to the impurity contribute to tunneling. While calculations based on the WKB approximation provide significant insight into the key band structure features determining the nature of BTBT in highly-mismatched alloys, quantitative analysis for the design of devices in which tunneling currents are of importance should employ a more rigorous Green's function-based approach that allows to compute the complex band structure self-consistently close in energy to localised states. Such models may also need to include contributions due to trap-assisted tunneling via defect states, \cite{Hurkx_IEEETED_1992} due to the presence of native defects in highly-mismatched alloys arising from the low growth temperatures required to promote N and Bi incorporation. \cite{Buyanova_JPCM_2004,Ludewig_SST_2015}


\section{Implications for practical applications}
\label{sec:implications}

As described in Sec.~\ref{sec:introduction}, for narrow-gap highly-mismatched III-V semiconductor alloys BTBT is most relevant to the operational characteristics of two classes of devices: (i) long-wavelength APDs, where BTBT contributes to leakage current, and (ii) TFETs, where BTBT provides the mechanism for current flow from the source into the channel. Having elucidated the impact of BAC in highly-mismatched alloys on BTBT, here we describe in turn the implications of our results for proposed practical applications in each of these classes of devices.


\subsection{Implications for APD applications}
\label{sec:implications_apd}

Our results in Sec.~\ref{sec:results_fixed_band_gap} highlight that BTBT in highly-mismatched alloys is governed by a field-dependent interplay between the impact of (i) the BAC-induced increase in CB or VB edge effective mass, which dominates at field strengths $\lesssim 1$ MV cm$^{-1}$ and acts to reduce the BTBT generation rate, and (ii) the BAC-induced increase in DOS close in energy to localised states, which dominates at field strengths $\gtrsim 1$ MV cm$^{-1}$ and acts to increase the BTBT generation rate. Electric field strengths corresponding to APD operation are typically in the range of several MV cm$^{-1}$, where our calculations suggest that the BTBT generation rate slightly exceeds that in a conventional III-V material of the same band gap. On this basis, it might be concluded that highly-mismatched alloys are not desirable for long-wavelength APD applications, as their band structure has the potential to produce higher leakage currents at fixed voltage. However, of primary concern for the development of long-wavelength APDs are $\alpha$ and $\beta$, the electron and hole impact ionisation rates (per unit length). From this perspective highly-mismatched alloys offer significant and as-yet little explored opportunities.

The potential benefit of employing highly-mismatched III-V alloys to develop long-wavelength APDs was proposed by Adams in Ref.~\onlinecite{Adams_EL_2004}. The excess noise factor of an APD is, at high multiplication, directly proportional to the ratio $k$ of $\alpha$ and $\beta$, with $k = \frac{ \beta }{ \alpha }$ or $\frac{ \alpha }{ \beta }$ for electron or hole multiplication respectively. To achieve low-noise multiplication therefore requires $k \ll 1$ -- i.e.~that multiplication of electrons (holes) via impact ionisation is, at fixed electric field, suppressed relative to multiplication of holes (electrons). \cite{David_IEEEJSTQE_2008} This is the case, e.g., in Si, where $\alpha \gg \beta$ so that photo-generated electrons predominantly initiate electron-hole pair generation via impact ionisation. However, $k \sim 1$ in most conventional III-V semiconductors, limiting the ability to achieve high signal-to-noise ratio at longer wavelengths via either electron or hole multiplication.

Adams proposed that this limitation could be overcome by exploiting the impact of N incorporation on the band structure of dilute nitride In(Ga)N$_{x}$As$_{1-x}$ -- where N incorporation strongly perturbs the CB structure, while leaving the VB structure comparatively unchanged -- to suppress electron multiplication. In In(Ga)N$_{x}$As$_{1-x}$, N incorporation results in a reduction of the electron mobility by approximately an order of magnitude -- e.g., from $\approx 8500$ cm$^{2}$ V$^{-1}$ s$^{-1}$ at room temperature in GaAs to $\lesssim 500$ cm$^{2}$ V$^{-1}$ s$^{-1}$ in GaN$_{x}$As$_{1-x}$ \cite{Mouillet_SSC_2003,Olea_PSSC_2010} -- due to a combination of (i) the BAC-induced increase in CB edge effective mass, and (ii) strong alloy scattering by N-related localised impurity states. \cite{Vaughan_PRB_2005,Vaughan_PRB_2007,Seifikar_PRB_2011} This strongly suppresses the electron drift velocity in response to an applied electric field, \cite{Seifikar_PRB_2011} so that N incorporation hinders electrons in acquiring sufficient energy to initiate carrier multiplication via impact ionisation. Dilute nitride alloys therefore present the opportunity to suppress $\alpha$ while leaving $\beta$ unchanged, and hence to achieve low-noise hole multiplication. However, initial experimental investigations suggested limited benefit to this approach, \cite{Tan_APL_2013} likely as it begins with an In(Ga)As semiconductor in which $\alpha > \beta$ and subsequently results, when N is incorporated, in a material which is closer to the unfavourable situation where $\alpha \approx \beta$.

Significantly more appealing is to exploit the same principle in dilute bismide In(Ga)As$_{1-x}$Bi$_{x}$ alloys, where Bi incorporation primarily impacts the VB structure while leaving the CB structure comparatively unchanged. Experimental investigations have demonstrated strong reduction in hole mobility in response to Bi incorporation \cite{Beaton_JAP_2010} -- with electronic structure calculations supporting suggestions based on experimental data of the presence of Bi-related localised impurity states that act as strong alloy scattering centres close in energy to the VB edge \cite{Usman_PRB_2011} -- while the electron mobility remains close to that in GaAs. \cite{Kini_JAP_2009} The dilute bismide case then mirrors the dilute nitride case: in a dilute bismide alloy one begins with a material in which $\alpha > \beta$, and further suppresses $\beta$ relative to $\alpha$ via Bi incorporation. This is a more desirable approach since it enables a reduced $k = \frac{ \beta }{ \alpha }$, with minimal reduction expected in the dominant electron multiplication rate, $\alpha$. Indeed, new experimental evidence suggests strong suppression of the hole impact ionisation rate in GaAs$_{1-x}$Bi$_{x}$ alloys, such that $\beta \ll \alpha$ with reported $k = \frac{ \beta }{ \alpha }$ values as low as 0.01. \cite{Liu_NC_2021}

Recent experimental work has demonstrated that prototypical telecom-wavelength APDs based on AlAs$_{1-x}$Sb$_{x}$ alloys offer significantly enhanced performance compared to conventional InP or Al$_{x}$In$_{1-x}$As devices, with high sensitivity operation demonstrated and impact ionisation rate ratio $k = \frac{ \beta }{ \alpha } \approx 0.005$ being significantly reduced compared to that in shorter wavelength Si APDs. \cite{Yi_NP_2019} Here, beginning from a conventional Al$_{x}$In$_{1-x}$As material, replacement of As by large, electropositive Sb atoms -- while simultaneously removing In to allow lattice matching to InP -- is expected to have a comparable effect as that suggested above for Bi-containing alloys. On this basis, we suggest that dilute bismide alloys are a promising material system for the development of low-noise III-V APDs, with the additional benefit that the large band gap reduction associated with Bi incorporation allows the operating wavelength to be tuned across a broad range in the near- and mid-infrared when grown on GaAs, \cite{Sweeney_JAP_2013,Broderick_IEEEJSTQE_2015} InP \cite{Jin_JAP_2013,Broderick_SST_2018} or InAs \cite{Webster_JAP_2016} substrates.


\subsection{Implications for TFET applications}
\label{sec:implications_tfet}

From the perspective of TFET applications, our results suggest three potential benefits of using narrow-gap highly-mismatched alloys as a channel material. Firstly, the ability to significantly alter the band gap $E_{g}$ via N or Bi composition provides a critical additional degree of freedom in the design of ultra-thin body (UTB) or gate-all-around (GAA) TFETs, allowing, e.g., to circumvent the reduction in current associated with the increased band gap induced by quantum confinement effects in conventional nm scale structures. Here, N or Bi can be incorporated in order to reduce the bulk band gap of InAs (cf.~Fig.~\ref{fig:alloy_band_gaps}), resulting in a reduced channel band gap in a UTB or GAA structure. Secondly, the suppression (enhancement) of the BTBT generation rate $G$ at low (high) applied electric field $F$ predicted above -- compared to that in a conventional (non-highly-mismatched) alloy having the same band gap -- indicates the possibility to achieve a higher ratio $\frac{ I_{\text{on}} }{ I_{\text{off}} }$ between the TFET ``on'' and ``off'' state currents $I_{\text{on}}$ and $I_{\text{off}}$. Thirdly, the enhanced difference between the low- and high-field BTBT generation rate in highly-mismatched alloys corresponds to an increased slope $\frac{ dG }{ dF }$ -- i.e.~a stronger increase in generation rate with increasing applied field compared to that in a conventional alloy having the same band gap. The BTBT generation rate is linked to the associated BTBT current density $J$ via $eGV = JF$, where $V$ is the applied voltage producing the effective electric field $F$ experienced by carriers in the channel region of a TFET device structure. The enhanced $\frac{ dG }{ dF }$ in highly-mismatched alloys can therefore be expected to lead to enhanced $\frac{ dJ }{ dV }$, and hence to increased subthreshold slope (reduced subthreshold swing).

Based on our calculated dependence of $G$ on $F$ at fixed band gap (cf.~Fig.~\ref{fig:4200_nm_transmission_generation}(d)), we note that the second and third benefits described above are expected to be more pronounced for InAs$_{1-x}$Bi$_{x}$ than for InN$_{x}$As$_{1-x}$. Atomistic quantum kinetic calculations for InAs TFETs suggest that while both GAA and UTB structures produce comparable $\frac{ I_{\text{on}} }{ I_{\text{off}} }$ as high as $10^{2}$, the two classes of TFET structure differ markedly in terms of achievable subthreshold swing.  \cite{Luisier_IEEEEDL_2009} Specifically, GAA structures offer significantly reduced subthreshold swing compared to single- or double-gate UTB structures, but while subthreshold swings as low as $28$ mV dec$^{-1}$ can be achieved in InAs GAA nanowire TFETs this is only possible when the channel (nanowire) diameter exceeds 10 nm. \cite{Luisier_IEEEEDL_2009} Here, incorporation of Bi to form an InAs$_{1-x}$Bi$_{x}$ channel in a GAA nanowire TFET offers clear benefits: the ability to reduce the band gap at fixed nanowire diameter is expected to relax constraints on the nanowire diameter required to achieve a $< 60$ mV dec$^{-1}$ subthreshold swing, allowing to achieve low subthreshold swing in a more compact geometry, while simultaneously allowing to increase the $\frac{ I_{\text{on}} }{ I_{\text{off}} }$ ratio due to the enhanced difference between the low- and high-field BTBT generation rate (cf.~Fig.~\ref{fig:4200_nm_transmission_generation}(d)).

On this basis, our analysis suggests that InAs$_{1-x}$Bi$_{x}$ alloys are a potentially promising material for the design of III-V TFETs displaying enhanced performance. In ideal (defect-free) structures, the performance of such devices is expected to be improved compared to that predicted for existing devices employing InAs as the channel material. However, it should be noted that the presence of native defects in highly-mismatched alloys is likely to generate deleterious trap-assisted BTBT, which has the potential to degrade the $\frac{ I_{\text{on}} }{ I_{\text{off}} }$ ratio and would be challenging to fully suppress in practice. Experimental investigations are therefore required in order to ascertain the degree to which the predicted enhanced BTBT characteristics of InAs$_{1-x}$Bi$_{x}$ alloys carry over to real devices, and hence the degree to which the performance of prototype InAs$_{1-x}$Bi$_{x}$-based TFETs can be enhanced over InAs-based TFETs.


\section{Conclusions}
\label{sec:conclusions}

In summary, we have presented a theoretical analysis of BTBT in narrow-gap, highly-mismatched III-V semiconductor alloys containing N and Bi. Specifically, by employing model BAC Hamiltonians derived and parametrised from atomistic alloy supercell electronic structure calculations, we analysed the impact of substitutional N or Bi impurities on the complex band structure, field-dependent BTBT transmission coefficient $T$, and field-dependent BTBT generation rate $G$. The impact of the impurity-like behaviour of substitutional N or Bi in InAs was emphasised via comparison of our calculations for highly-mismatched InN$_{x}$As$_{1-x}$ and InAs$_{1-x}$Bi$_{x}$ to those for the conventional alloy InAs$_{1-x}$Sb$_{x}$. Given the strong dependence of $T$ and $G$ on the band gap, the impact of BAC was firstly quantified by comparing field-dependent $T$ and $G$ for these alloys at fixed band gap, before trends as a function of alloy band gap were described.

Our analysis reveals that, at fixed band gap, the impact of BAC on BTBT is dominated by a field-dependent competition between the increased band edge effective mass (at low field), and the associated increased DOS close in energy to the N- or Bi-related impurity state (at high field). InN$_{x}$As$_{1-x}$ and InAs$_{1-x}$Bi$_{x}$ are alloys in which BAC effects are respectively weak and strong, due to the separation in energy between the relevant host matrix band edge and localised impurity state in each case. Correspondingly, Bi incorporation in InAs was observed to more strongly impact BTBT than N incorporation. At low field $F \lesssim 1$ MV cm$^{-1}$, the increased CB or VB edge effective mass increases the area bounded in energy and wave vector by the complex band linking the VB and CB, reducing $G$ compared to that in InAs$_{1-x}$Sb$_{x}$, by $\approx 20$\% and $50$\% in InN$_{x}$As$_{1-x}$ and InAs$_{1-x}$Bi$_{x}$ respectively. At high field $F \gtrsim 1$ MV cm$^{-1}$, the increased DOS close in energy to localised impurity states limits the growth of this area, reflecting that more states are available to contribute to BTBT in a given energy range and acting to increase $G$ above that in InAs$_{1-x}$Sb$_{x}$.

From the perspective of device applications, our calculations indicate that APDs based on highly-mismatched alloys are unlikely to display reduced leakage currents compared to those based on conventional alloys, as $G$ in the relevant range of operating electric field strengths ($\sim$ MV cm$^{-1}$) is comparable to that in conventional alloys. However, the expected strong reduction in hole impact ionisation rate in dilute bismide alloys is highly promising, making dilute bismide alloys of continuing interest for low-noise, long-wavelength APD applications. Our calculations further indicate that highly-mismatched alloys possess greater contrast in low vs.~high field BTBT generation rate compared to conventional alloys. This is promising from the perspective of TFET applications, suggesting a novel route to engineer the electrical characteristics via the complex band structure of the channel, and hence providing an new approach to employ in the design of TFETs displaying enhanced subthreshold slope.

Investigations of the impact of BAC in highly-mismatched alloys have to date centred largely on modifications to the real band structure and their implications for applications in photonic or photovoltaic devices. Our analysis provides new insight into the properties of this interesting class of semiconductors by revealing that BAC interactions provide scope to modify the complex band structure and BTBT, thereby providing new opportunities for device applications.


\section*{Acknowledgements}

This work was supported by Science Foundation Ireland (SFI; project no.~15/IA/3082), and by the National University of Ireland (NUI; via the Post-Doctoral Fellowship in the Sciences, held by C.A.B.). The authors thank Mr.~Michael D.~Dunne (Tyndall National Institute, Ireland) for useful discussions.



%

\end{document}